\documentstyle[aps,multicol]{revtex}
\renewcommand{\narrowtext}{\begin{multicols}{2} \global\columnwidth20.5pc}
\renewcommand{\widetext}{\end{multicols} \global\columnwidth42.5pc}
\input{epsf.tex}
\def\inseps#1#2{\def\epsfsize##1##2{#2##1} \centerline{\epsfbox{#1}}}

\def\top#1{\vskip #1\begin{picture}(290,80)(80,500)\thinlines \put(
65,500){\line( 1, 0){255}}\put(320,500){\line( 0, 1){
5}}\end{picture}}
\def\bottom#1{\vskip #1\begin{picture}(290,80)(80,500)\thinlines \put(
330,500){\line( 1, 0){255}}\put(330,500){\line( 0, -1){
5}}\end{picture}}

\def\Btau{\bbox{\tau}}

\def\jmag{{\bbox J_{\rm mag}}}
\def\jtr{{\bbox J_{\rm tr}}}
\def\r{{\bbox r}}
\def\rp{{(\bbox r)}}
\def\nab{{\bbox \nabla}}
\def\F{{\bbox F}}
\def\E{{\bbox E}}
\def\p{{\bbox p}}
\def\A{{\bbox  A}}
\def\j{{\bbox j}}
\def\q{{\bbox q}}
\def\B{{\bbox  B}}
\def\M{{\bbox M}}
\def\J{{\bbox J}}

\def\v{{\bbox v}}
\def\q{{\bbox q}}
\def\hx{{\hat{\bbox{x}}}}
\def\l1{{\hat L^{(1)}}}

\def\S{{\bbox S}}
\def\I{{\bbox I}}
\def\hn{{\hat {\bbox{n}}}}
\def\hz{{\hat {\bbox{z}}}}
\def\bulk{{\rm bulk}}
\def\hep{{\hat {\bbox{\epsilon}}}}

\begin{document}
\draft

\title{Thermoelectric Response of an Interacting Two-Dimensional 
Electron Gas in Quantizing Magnetic Field}
\author{N.R. Cooper\cite{ill}, B.I. Halperin, and I.M. Ruzin\cite{dmm}}
\address{Lyman Laboratory of Physics, Harvard University,
Cambridge, MA 02138.}
\date{\today}

\maketitle

\begin{abstract}
We present a discussion of the linear thermoelectric response of an
interacting electron gas in a quantizing magnetic field. Boundary
currents can carry a significant fraction of the net current passing
through the system. We derive general expressions for the bulk and
boundary components of the number and energy currents. We show that
the local current density may be described in terms of ``transport''
and ``internal magnetization'' contributions. The latter carry no net
current and are not observable in standard transport experiments. We
show that although Onsager relations cannot be applied to the local
current, they are valid for the transport currents and hence for the
currents observed in standard transport experiments.  We relate three
of the four thermoelectric response coefficients of a disorder-free
interacting two-dimensional electron gas to equilibrium thermodynamic
quantities. In particular, we show that the diffusion thermopower is
proportional to the entropy per particle, and we compare this result
with recent experimental observations.
\end{abstract}


\narrowtext

\section{INTRODUCTION}

It has long been realized that surface (boundary) currents provide a
significant contribution to the thermoelectric response of electronic
systems in a quantizing magnetic field. The first calculations taking
boundary currents into account were presented by Obraztsov\cite{obr65}
for non-interacting electrons. Such issues have been important in
recent theories of the thermoelectric properties of two-dimensional
systems in the integer quantum Hall regime. Calculations of the
thermoelectric response of non-interacting electrons in this regime
have been presented by a number of authors using various
approaches\cite{girvinjonson,streda,jon,oji,oji85,gru}.  Measurements
of the intrinsic, ``diffusion-and-drift thermopower'' in the integer
quantum Hall regime are consistent with theories for non-interacting
electrons, once disorder is introduced. (For a review of experiments
and theories of the intrinsic and phonon-drag thermopower for
non-interacting electrons, see Ref.~\onlinecite{gal}.)

Recently, however, there have been reports of measurements of the
diffusion thermopower in the fractional quantum Hall
regime\cite{ying,bayot}.  It is clear, both from these measurements
and from what is known of the fractional quantum Hall effect, that
interactions must play an important role in determining the transport
properties in this regime.

In this paper we discuss the thermoelectric response of an interacting
electron gas, paying particular attention to the importance of the
boundary currents. In section~II we restate the general expressions
for the linear response, following an approach first proposed by
Luttinger\cite{lut}, and since discussed for the quantum Hall regime
by Oji and Streda\cite{oji85}. We extend this analysis by deriving
general expressions for the local energy-current and number-current
distributions in gradients of temperature and chemical potential. We
argue that, even when electron-electron interactions are included,
both the number and energy currents in the bulk of the sample may be
separated into ``transport'' and ``internal magnetization''
contributions.  The magnetization contribution causes no net current
to flow through the sample. However, it can have a significant effect
on the local current density. We show that Onsager relations may still
be applied, in a quantizing magnetic field, for the transport
currents, and hence for the net currents through the sample. (The
Onsager relations cannot, in general, be applied directly to the local
current densities.)

In Section~III, we consider a two-dimensional electron system in the
limit of zero impurity scattering, and we derive the forms of various
transport coefficients in this case.  In particular, the thermopower
coefficient is shown to be equal to the entropy per carrier divided by
the charge of the carrier, a result first derived by Obraztsov for
non-interacting electrons.  In Section~IV, we compare these results
with recent data, at very low temperatures, on $p$-type samples with
Landau-level filling fractions near $\nu=1/2$ and
$\nu=3/2$\cite{ying}.  We find that the data at $\nu=1/2$ are
consistent with an interpretation based on a model of spin-polarized
``composite fermions'', with a reasonable value of the effective mass,
but this does not seem to be the case at $\nu=3/2$.

Many of the results of Section~II, particularly for the net
currents, have been obtained previously by Oji and Streda\cite{oji85},
at least for the case where the gradients of the potentials and the
temperature are constant throughout the sample. Many details were
omitted from their presentation, however, and many of the underlying
assumptions were not stated explicitly. Because there are a number of
subtle points in the derivation, because there appears to have been
some confusion in the literature\cite{wu}, and because the results are
of fundamental importance, we give here a detailed and general
derivation.

\section{GENERAL EXPRESSIONS FOR LINEAR RESPONSE}

\subsection{General Considerations}
\label{subsec:gencon}

We begin by reviewing the ``hydrodynamic'' assumptions inherent in any
theoretical discussion of transport coefficients such as the thermal
and electrical conductivities, thermopower, etc.  We restrict our
attention to small deviations from thermal equilibrium, in samples
which are very large compared to atomic distances or other microscopic
length scales, and we shall investigate the response to weak external
perturbations which vary slowly in space and in time.

We assume here that particles interact only via short-range forces,
deferring until subsection~G the modifications necessary in the
presence of long-range Coulomb interactions.

The fundamental hydrodynamic assumption is that there exists a
microscopic relaxation rate $\tau_m$, such that for perturbations
which vary on a time scale slow compared to $\tau_m$, the system
relaxes to a state that is close to local thermodynamic equilibrium,
and where all properties of interest may be described in terms of an
expansion about local equilibrium.  More particularly, one identifies
a set of conserved quantities, which in the systems of interest to us
are the energy $E$ and particle number $N$, and one defines
corresponding conserved densities, such as the energy density
$\epsilon(\r)$ and particle density $n (\r)$. On time scales large
compared to $\tau_m$, one assumes that all physical quantities
localized near a point $\r$ relax to values which are determined by
the values of the conserved densities and their low-order spatial
derivatives in the vicinity of $\r$. On the other hand, one cannot in
general assume that the conserved quantities themselves relax to their
equilibrium values in a microscopic time scale. The conservation laws
relate the time derivatives of conserved quantities to the divergence
of associated transport currents, and these time derivatives may be
very small if the length scale of the system is large.

In systems with short-range forces, the slowest relaxations are
typically characterized by a diffusion coefficient $D$, so that the
slowest relaxation time for the conserved densities is given by
$\tau_M\approx L^2/D$, where $L$ is either the size of the system or
the wavelength of the perturbation, whichever is shorter. Clearly if
$L$ is very large, $\tau_M$ may be very much larger than $\tau_m$.
Although the overall response to an external perturbation may be quite
different in the limits where the frequency is large or small compared
to $\tau_M^{-1}$, the hydrodynamic equations themselves are assumed to
apply for time scales $\tau$ large compared to $\tau_m$, regardless of
whether $\tau$ is large or small compared to $\tau_M$.

Our central focus will be on the particle current density $\J(\r)$ and
the energy current density $\J^E(\r)$.  At least in cases where there
are only short-range interactions between particles, the hydrodynamic
assumptions imply, in particular, a {\it locality hypothesis} for $\J$
and $\J^E$, viz. that $\J$ and $\J^E$ are determined by the values of
$\epsilon(\r)$ and $n(\r)$ and their variations, only in the immediate
vicinity of point $\r$. The currents may also depend on {\it local
material parameters} such as the local chemical composition, impurity
concentration, etc., and on the applied magnetic field $\B$. We assume
the material parameters to be independent of time, but they may depend
on position in cases of interest.  If the material parameters are
independent of position, then the locality hypothesis implies that for
variations on a time scale slow compared to the microscopic relaxation
rate $\tau_m^{-1}$, the current densities $\J$ and $\J^E$ at point
$\r$ may be considered to be functions of $\epsilon$, $n$, and their
gradients at point $\r$.

The locality assumption, central to any hydrodynamic description, is
difficult or impossible to prove rigorously under general conditions.
One important piece of the physics, which enters the case of quantum
systems in a strong magnetic field, is that a sample can have non-zero
``magnetization currents'' even in a situation of thermodynamic
equilibrium. In this case, a proof that the electron number current
$\J(\r)$ is independent of conditions far from $\r$ is equivalent to
proving that the equilibrium magnetization $\M(\r)$ (defined below)
depends only on local conditions. For a small sample, $\M(\r)$ may in
fact be rather sensitive to conditions at relatively large distances
from $\r$. For example, for a small metallic loop, at very low
temperatures, the equilibrium current depends on the magnetic flux
through the loop, modulo units of the flux quantum $hc/e$, because of
the Aharonov-Bohm effect. Such effects, however, become negligible in
the ``thermodynamic'' limit of large sample sizes. Since the
equilibrium magnetization density can be related by thermodynamics to
a derivative of the free energy with respect to the applied magnetic
field [cf.  Eq.~(\ref{eq:B43}) below] a proof of the locality of $\M$
reduces to proving that a system has a well-behaved thermodynamic
limit for the free energy, at any non-zero temperature.

In our analysis, it will be convenient to eliminate $n(\r)$ and
$\epsilon(\r)$, in favor of two suitably defined ``statistical''
fields: a local chemical potential $\mu(\r)$ and a local temperature
$T(\r)$. We shall also introduce shortly, external ``mechanical
fields'': a potential $\phi(\r)$ which couples to the number density,
and a fictitious ``gravitational potential'' $\psi(\r)$ which couples
to the energy density. We also define an electrochemical potential
$\xi=\mu+\phi$.  We shall see that in thermodynamic equilibrium (even
if the material parameters vary in space) $\xi$ and $(1+\psi)T$ are
constants in space.

In cases where there is time-reversal symmetry (hence $B=0$), the
currents $\J^E$ and $\J$ must vanish in thermodynamic equilibrium.
Therefore, the first terms in the gradient expansions for $\J^E$ and
$\J$ must be proportional to $\nab\xi$ and $\nab T$ (assuming
$\psi=0)$.  For a quantum mechanical system in the presence of an
applied magnetic field, however, there may be non-zero circulating
currents even in a situation of thermodynamic equilibrium, as was
noted above.  We shall find it convenient to break the currents $\J$
and $\J^E$ into a ``transport'' part and a ``magnetization'' part
according to
\begin{eqnarray}
\J(\r) & = & \jtr(\r)+ \jmag(\r),
\label{eq:B1}
\\
\J^E(\r) & = &  \J_{\rm tr}^E(\r)+ \J_{\rm mag}^E(\r),
\label{eq:B2}
\end{eqnarray}
where $\jtr$ and $\J_{\rm tr}^E$ vanish in thermodynamic 
equilibrium and 
\begin{eqnarray}
\jmag(\r) & = & \nab\times\M^N(\r),
\label{eq:B3}
\\
\J_{{\rm mag}}^E(\r) & = & \nab\times\M^E(\r).
\label{eq:B4}
\end{eqnarray}
The ``magnetization densities'' $\M^N\rp$ and $\M^E\rp$ are defined to
be functions of the temperature and chemical potential only at the
given point $\r$. These functions, in turn, may be computed in thermal
equilibrium; i.e., we may compute the values of $\M^N\rp$ and
$\M^E\rp$ assuming that $\mu$ and $T$ are independent of position.
(Any applied ``mechanical potential'' may also be taken independent of
$\r$.)

We make a number of observations about the magnetization currents.

1. If we consider a homogeneous sample with sharp boundaries, in
thermodynamic equilibrium, then the magnetizations $\M^N$ and $\M^E$
are uniform and the magnetization currents vanish in the interior of
the sample. However, there will in general be currents flowing on the
surface of the sample. As one can readily derive from
Eqs.~(\ref{eq:B3}) and~(\ref{eq:B4}), the surface current densities at
a point $\Btau$ on the boundary are given by
\begin{eqnarray}
\I & = & \M^N\times\hn,
\label{eq:B5}
\\
\I^E & = & \M^E\times\hn,
\label{eq:B6}
\end{eqnarray}
where $\hn$ is the unit vector normal to the surface, pointing outward
from the sample. The same expressions for the boundary currents are
also valid for an inhomogeneous sample, in which case the
magnetizations $\M^N, \M^E$ vary in the sample. Their values in
Eqs.~(\ref{eq:B5})~(\ref{eq:B6}) should be determined at a point
inside the boundary close to $\bbox{\tau}$.

We are assuming here, and throughout this paper, that the material
exterior to the sample is either a vacuum or an ``ideal'' material
with no magnetization of its own. Otherwise, we would have a second
contribution to the edge current from the magnetization of the
exterior medium.

2. For a two-dimensional conducting layer in a semiconductor system,
the magnetizations $\M^N$ and $\M^E$ are normal to the layer. If the
magnetization lies in the positive $\hz$ direction, then the boundary
currents will point parallel to the sample edge in the
counter-clockwise direction, looking down at the sample.

3. The integrated boundary currents given by Eqs.~(\ref{eq:B5})
and~(\ref{eq:B6}) depend on the magnetization at a point just inside
the sample, but are independent of such details as whether the
boundary is sharp or diffuse on the atomic scale, the concentration of
impurities near the boundary, etc.  In the case of thermodynamic
equilibrium, in a uniform sample, where the magnetizations are
independent of position, the surface currents are divergence-free.
This is of course necessary since there should be no bulk currents in
this case. Alternatively, we see that the condition of divergence-free
surface currents, (\ref{eq:B5}), (\ref{eq:B6}), for arbitrary surface
treatments, requires that the magnetizations are truly bulk
properties, independent of any details of the surface.

4. In our discussion of electron systems, the only {\it particles}
which are allowed to move over macroscopic distances are the
conduction electrons.  The electrical current $\J^e$ is related to the
particle current $\J$ of the electrons by $\J^e=-e\J$, where $(-e)$ is
the electron charge.  For a sample at equilibrium the magnetization
$\M^N$ is related to the conventional magnetic moment density $\M$, in
Gaussian units, by
\begin{equation}
\M=(-e/c)\M^N.
\label{eq:B7}
\end{equation}

To avoid confusion, we note, that the quantity $\M$, defined by
Eqs.~(\ref{eq:B1}) and~(\ref{eq:B7}), has a direct physical meaning in
terms of the magnetic moment per unit volume only if the sample is at
equilibrium. In the non-equilibrium case, the total magnetic moment
$\bbox{\cal M}$ is determined by the total current density distribution
$\J$, including both the ``transport'' component $\J_{\rm tr}$ and the
``magnetization'' component $\J_{\rm mag}$, as given by the general
formula
\begin{equation}
\bbox{\cal M}=\frac{-e}{c}\int{d^3\bbox{r}[\r\times (\J(\r)-\overline\J)]},
\end{equation}
where the overline denotes the volume average.  Moreover, since
$\J_{\rm tr}$ cannot, in general, be expressed as the curl of a vector
field, the magnetic moment of the sample in the non-equilibrium case
cannot be expressed as an integral of any local magnetization density.

5. For a uniform macroscopic sample, in thermal equilibrium, the
magnetization may be related to other thermodynamic quantities. By
definition, the magnetization $\M$ is equal to $-(1/V)dE/d\B$, when
the entropy $S$, the volume $V$, and the electron number $N$ are held
fixed. More generally, we may write
\begin{equation}
TdS=dE+PdV+\M V\cdot d\B-\mu dN, 
\label{eq:B43}
\end{equation}
where $\mu$ is the chemical potential of the electrons, 
and $P$ is the electron ``pressure''. (We are assuming 
here short-range forces between the electrons.) From the 
extensivity properties of $S$, $E$ and $N$, it then 
follows that
\begin{equation}
n\mu=\epsilon-Ts+P,
\label{eq:B44}
\end{equation}
where $\epsilon$ and $s$ are the energy and entropy per 
unit volume. Then using (\ref{eq:B7}) we find
\begin{eqnarray}
 & & \M^N  =  -{c\over e}\left.{\partial P\over \partial \B}
\right|_{\mu,T}, 
\label{eq:B45}
\\
 & & P =  \int_{-\infty}^\mu
n(\mu',T,\B)\;d\mu'. 
\label{eq:B46}
\end{eqnarray}

6. If there are temperature or electrochemical potential gradients,
there may be non-vanishing magnetization currents in the interior of
an otherwise uniform sample. Alternatively, if there are gradients in
the material parameters, bulk magnetization currents can be present
even at thermodynamic equilibrium.

7. Under all conditions, the magnetization currents $\jmag$ and
$\J_{\rm mag}^E$ are divergence-free. As a consequence, these currents
do not make any contribution to the {\it net} current flows that are
measured by conventional transport experiments.  Specifically,
consider any closed curve $C$ that encircles the sample but is
exterior to it, and let $S$ be a surface spanning this contour. The
total magnetization currents crossing the surface must be zero, by
Stokes' theorem, and the total currents $I_S$ and $I_S^E$ crossing the
surface are obtained by considering the transport currents alone:
\begin{eqnarray}
I_S & \equiv & \int_S\hn\cdot
\J\;d^2S=\int_S\hn\cdot\jtr\;d^2S ,
\label{eq:B10}
\\
I_S^E & \equiv & \int_S\hn\cdot
\J^E\;d^2S=\int_S\hn\cdot\jtr^E\;d^2S ,
\label{eq:B11}
\end{eqnarray}
where $\hn$ is the local normal to $S$. In a d.c. transport
experiment, where $\nab\cdot\J=\nab\cdot\J^E=0$, the currents $I_S$
and $I_S^E$ will be independent of the particular surface chosen.
This argument applies equally well to a singly- or multiply-connected
sample including the case where $C$ threads a hole in the sample, as
in Figure~1. In the case of a two-dimensional sample, the surface $S$
becomes a curve traversing the sample, while $I_S$ and $I_S^E$ are the
total currents across the curve.
\begin{figure}
\inseps{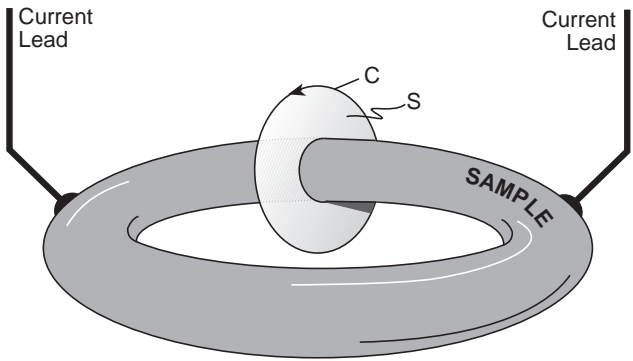}{0.6}
\vskip0.35cm
\caption{Contour C, external to sample, is spanned by surface $S$. The
total current through $S$ is equal to the transport current across
$S$, as the magnetization current gives no contribution. The
illustration shows a multiply-connected sample.}
\end{figure}

In the remainder of this section, we shall apply the above
considerations to the general hydrodynamic description of an electron
system in a strong magnetic field. Our goal is to obtain some
relations among the various transport coefficients -- both for the
transport currents $\J_{{\rm tr}}\rp$ and $\J_{{\rm tr}}^E\rp$ and for
the total local currents, $\J\rp$ and $\J^E\rp$. We also relate the
transport coefficients to microscopic expressions involving
correlation functions for currents in the equilibrium state.

Our strategy can be best illustrated by considering a simple example
where the electrons are subject to a weak external potential
$\phi(\r)$, while the temperature is maintained constant, say by
contact with a substrate. Then inside the sample we may write
\begin{equation}
\J_{{\rm tr}}\rp=-\hat N^{(1)}
\nab\mu\rp-\hat L^{(1)}\nab\phi\rp ,
\label{eq:mynum}
\end{equation}
where $\mu\rp$ is the local chemical potential for the electrons, and
$\hat N^{(1)}$ and $\hat L^{(1)}$ are second-rank tensors, which we
denote ``transport coefficients.'' In thermodynamic equilibrium, the
electrochemical potential $\xi(\r)=\mu(\r)+\phi(\r)$ is independent of
position, and $\J_{{\rm tr}}$, by definition is zero.  Since this must
be true for arbitrary $\phi\rp$, we immediately obtain the ``Einstein
relation'' $\hat N^{(1)}=\hat L^{(1)}$.

Macroscopic equations for the response of the system to a
time-dependent perturbation $\phi(\r,t)$ are obtained by combining
(\ref{eq:mynum}) with the conservation law $(\partial n/\partial
t)=-\nab\cdot\jtr$. (Recall that $\nab\cdot \J_{{\rm mag}}=0$.)  If
there is a periodic disturbance in $\xi\rp$ with a wave vector
$\q||\hx$, then the electron density will relax towards the
equilibrium state, with $\xi(\r)=$~const, at a rate $\gamma_q=Dq^2$,
where the diffusion constant $D$ is given by
\begin{equation}
D= L_{xx}^{(1)}/(\partial\mu/\partial n).
\end{equation}
(Again, we assume short-range interactions between the electrons, so
$\partial\mu/\partial n$ is finite for $q\rightarrow 0$.)  A
microscopic expression for $\l1$ can be obtained by using quantum
mechanics to calculate the response of the system to an infinitesimal
time-dependent perturbation $\phi$, applied at a frequency $\omega$
which is small compared to microscopic frequencies $\tau_m^{-1}$, but
high compared to $\gamma_q$, so that the density $n$ does not have
time to change significantly. One thus obtains an expression for $\l1$
in terms of a two-time correlation function for fluctuations in $\J$
in the equilibrium state.

In order to generalize this procedure to the case of non-uniform
temperature, we follow the work of Luttinger\cite{lut} which involves
the introduction of a fictitious ``gravitational potential''
$\psi\rp$, which enters the Hamiltonian through its coupling to the
energy density. If linear response to $\psi$ is calculated, the
response to a temperature gradient may be obtained from the Einstein
relations.  Our situation is more complicated than the case considered
by Luttinger, however, because the Einstein relations apply only to
the transport currents, and not to the total currents.

\subsection{Current Operators in the Presence
of Electrical and Gravitational Fields}

The linear response to mechanical fields (electrostatic and
gravitational) in a quantizing magnetic field has been presented
previously\cite{oji85}. However, since it is important to the rest of
our discussion, we shall review this here.  We consider a Hamiltonian
of the form
\widetext
\begin{equation}
H_0 \equiv  \sum_{i=1}^N h_i
 =  \sum_{i=1}^N  
\left\{ \frac{\left[ \p_i+e\A(\r_i)/c \right]^2}{2m}+V(\r_i)+{1\over 2}
\sum_{j (\neq i) =1}^N u_{ij} \right\} ,
\label{eq:1}
\end{equation}
\bottom{-2.7cm}
\narrowtext
\noindent
where $\A(\r)$ is the magnetic vector potential, $V(\r)$ is the scalar
potential energy including the confinement and disorder potentials and
the periodic potential of atomic cores, and $u_{ij}$ describes the
interparticle interactions. Following Luttinger, we introduce the
number and energy density operators\cite{lut,mor}
\begin{eqnarray}
 & & \rho\rp  \equiv  \sum_i\delta_i , 
\label{eq:2}  \\
 & & h\rp  \equiv  {1\over 2}\sum_i \{ h_i, \delta_i \}, 
\label{eq:3}
\end{eqnarray}
where $\delta_i\equiv \delta (\r-\r_i)$, $h_i$ is defined in equation
(\ref{eq:1}), and the curly brackets indicate the anticommutator, $ \{
A,B \} \equiv AB+BA$.

We are interested in the response of the current-density to time
varying external ``electrostatic'' and ``gravitational'' potentials
$\phi$ and $\psi$, respectively. These fields couple to the number and
energy densities according to the Hamiltonian
\begin{eqnarray}
& & H_T  =  \int \! d^3r\; h_T\rp, 
\label{eq:6a}
\\
& & h_T\rp  \equiv  h\rp+\phi\rp\rho\rp+
\psi\rp h\rp . 
\label{eq:6b}
\end{eqnarray}
[We call the function $\psi\rp$ defined by Eqs.~(\ref{eq:6a})
and~(\ref{eq:6b}) the ``gravitational potential'' to follow the
terminology of the original work by Luttinger\cite{lut}, although the
true gravitational potential would also be coupled to the mass density
rather than just to the first-order term of the relativistic energy
expansion.]

The conservation laws for energy and particle number imply that the
Heisenberg equations of motion for $\rho$ and $h_T$, under the
Hamiltonian $H_T$, may be written in the form\cite{lut}
\begin{eqnarray}
{d\rho\rp\over dt} & = & - {i\over\hbar} [\rho, H_T]= -\nab\cdot\J\rp ,
\label{eq:B13}
\\
{dh_T(\r)\over dt} &  = & - {i\over\hbar}[h_T,H_T]+{\partial
h_T\over\partial t}= -\nab\cdot\J^E\rp+ {\partial h_T\rp\over\partial
t},
\label{eq:B14}
\end{eqnarray}
where $\J$ and $\J^E$ are here operators for the particle and energy
currents. The second term on the right-hand side of (\ref{eq:B14})
occurs when there is a time dependence to $\phi\rp$ or $\psi\rp$. The
above equations constrain, but do not determine uniquely, the forms of
the operators $\J$ and $\J^E$, as we may in principle add to them an
arbitrary function whose divergence is identically zero. A requirement
of strict locality for the particle current, however, together with
Eqs.~(\ref{eq:6a}) and~(\ref{eq:6b}), imposes the form
\begin{equation}
\J(\r)=\j\rp[1+\psi\rp],
\label{eq:B15}
\end{equation}
where $\j\rp$ is the particle current for $\psi=0$:
\begin{eqnarray}
& & \j(\r) \equiv  
{1\over 2}\sum_i\{\v_i,\delta_i\} , 
\label{eq:B16}
\\
& & \v_i \equiv  [\p_i+e\A(\r_i)/c]/m. 
\label{eq:B17}
\end{eqnarray}

The requirement of strict locality cannot be applied either to the
energy current or the energy density in a non-relativistic theory with
interactions of non-zero range. However, the form of the energy
current is adequately restricted for our purposes if we require that
it depends only on the positions and velocities of particles in a
small neighborhood of $\r$.  Although various forms of the energy
current may still be written down, which are consistent with this
requirement and with a requirement that Eq.~(\ref{eq:B14}) be
satisfied exactly on the microscopic scale, we shall adopt here the
particular definitions
\widetext
\top{-2.8cm}
\begin{eqnarray}
& & \J^E(r) \equiv 
\j^E\rp+
\phi\rp\j\rp+2\psi\rp \j^E\rp, 
\label{eq:B18}
\\
& & j_\alpha^E\rp \equiv  
{1\over 4}\sum_i\{h_i,\{v_i^\alpha,\delta_i\}\} +
{1\over 8}\sum_{i\not= j}\sum_\gamma
\{(v_i^\gamma+v_j^\gamma),\tau_{ij}^{\alpha\gamma}\rp\} , 
\label{eq:B19}
\\
& &
\tau_{ij}^{\alpha\gamma}\rp  \equiv 
(r_i^\alpha-r_j^\alpha)\;F_{ij}^\gamma
\;\Delta_{ij}\rp , 
\label{eq:tau}
\\
& &
\Delta_{ij}\rp \equiv 
\int_0^1\!\!\!\!ds\;\delta[\r-\r_j-(\r_i-\r_j)s] , 
\label{eq:21}
\end{eqnarray}
\bottom{-2.7cm}
\narrowtext
\noindent
where $F_{ij}^\alpha\equiv-\partial u_{ij}/\partial r_i^\alpha$ is the
force on particle $i$ due to its interaction with $j$. We note that by
definition $\j^E$ is the energy current associated with the
unperturbed Hamiltonian $H_0$, and equation (\ref{eq:B18}) is valid to
first order in $\psi$, with the requirement that $\phi$ and $\psi$
vary very slowly in space compared to the range of the interaction
$u_{ij}$.  It is not difficult to show, under these circumstances,
that $\J^E$ exactly satisfies the Heisenberg equation of motion
(\ref{eq:B14}).  To demonstrate this, it is convenient to integrate
Eq.~(\ref{eq:B14}) over a small volume $\delta V$, and to write the
integral of $\nab\cdot \J^E$ as an integral over the surface $S$ of
the volume.  We also note that
\begin{equation}
\int_S\Delta_{ij}
\rp\;(\r_i-\r_j)\cdot d\S=\chi_j-\chi_i,
\label{eq:chi}
\end{equation}
where $\chi_i\equiv 1$, if particle $i$ is inside the volume enclosed
by $S$, and zero otherwise.

With our definitions, the contribution to the energy current from the
interaction potential $u_{ij}$ is distributed along the straight-line
segment joining $\r_i$ and $\r_j$, similar to the distribution of the
energy current adopted in Ref.~\onlinecite{forster}. By contrast, in
Ref.~\onlinecite{lut}, Luttinger employed a different form for $\j^E$,
where the energy current is supposed to be localized at the two points
$\r_i$ and $\r_j$.  Luttinger's expression is only approximate on the
microscopic level, since it does not strictly satisfy
Eq.~(\ref{eq:B14}).  Our more precise formula and Luttinger's are
equivalent, however, when integrated over any volume large compared to
the range of $u_{ij}$. In fact, in the present paper we have no real
need for a precise form for $\j^E(\r)$, but will only make essential
use of (\ref{eq:B18}).

\subsection{Thermodynamic Equilibrium}

We first consider the situation of thermodynamic equilibrium.  A
grand-canonical ensemble, for the perturbed Hamiltonian $H_T$ may be
described by a density matrix
\begin{equation}
w=Z^{-1}e^{-\beta(H_T-\xi N)},
\label{eq:B201}
\end{equation}
where $\xi$ and $\beta\equiv T_0^{-1}$ are Lagrange multipliers, which
we may describe, respectively, as the ``electrochemical potential''
and the ``inverse thermodynamic temperature.''

The properties of $w$ are particularly simple in the case where $\psi$
and $\phi$ are constants, independent of position.  Then the
eigenstates of $H_T$ are the same as the eigenstates of $H_0$, and the
density matrix $w$ is identical to a density matrix $w_0$, of the form
\begin{equation}
w_0=Z^{-1}e^{-(H_0-\mu N)/T},
\label{eq:B202}
\end{equation}
with
\begin{eqnarray}
& & T^{-1}  =  \beta(1+\psi), 
\label{eq:B203}
\\
& & \mu = (\xi-\phi)/(1+\psi). 
\label{eq:B204}
\end{eqnarray}
We denote $T$ as the ``internal temperature,'' and $\mu$ as the
``internal chemical potential''. The internal energy density
$\epsilon\rp\equiv\langle h\rp\rangle$ and the particle density
$n\rp\equiv\langle\rho\rp\rangle$ are the same in the two cases, as is
the entropy $S\equiv -{\rm tr} (w\ln w)$. Note that $\epsilon$ and $n$
may depend on position, even in equilibrium, if the materials
parameters vary in space.

The particle and energy currents are not the same in the two cases,
however, because the operators themselves are modified by the presence
of $\psi$ and $\phi$. Using (\ref{eq:B15}) and (\ref{eq:B18}), we see
that
\begin{eqnarray}
\langle\J\rp\rangle & = &
\langle\j\rp\rangle+
\psi\langle \j\rp\rangle,
\label{eq:B205}
\\
\langle\J^E\rp\rangle & = &
\langle \j^E\rp\rangle+
\phi\langle \j\rp\rangle+
2\psi\langle \j^E\rp\rangle.
\label{eq:B206}
\end{eqnarray}
The expectation values on the right-hand side represent the particle
and energy currents in the ``unperturbed'' state, with $\phi=\psi=0$,
temperature $T$, and chemical potential $\mu$. [Note: in
Eq.~(\ref{eq:B206}), as in Eq.~(\ref{eq:B18}) we have dropped terms of
order $\psi^2$, since we always consider $\psi$ to be small.]

The currents in Eqs.~(\ref{eq:B205}) and~(\ref{eq:B206}) are purely
magnetization currents, since the transport currents vanish, by
definition, in equilibrium. In the case of a uniform sample, the
magnetization currents are confined to the boundaries.  More
generally, they are given by Eqs.~(\ref{eq:B3}) and~(\ref{eq:B4}).
This implies that in the equilibrium state, with constant $\phi$ and
$\psi$, we have, to first order in the perturbations, at any point
$\r$ in the sample
\begin{eqnarray}
\M^N & = & (1+\psi)\M_0^N(\mu,T),
\label{eq:B20}
\\
\M^E & = & (1+2\psi)\M_0^E(\mu,T)+
\phi \M_0^N(\mu,T),
\label{eq:B21}
\end{eqnarray}
where $\M_0^N$ and $\M_0^E$ are the unperturbed magnetizations,
(corresponding to $\psi=\phi=0$), which we have written explicitly as
functions of the chemical potential $\mu$ and internal temperature
$T$. (The functions $\M_0^N$ and $\M_0^E$ are also implicitly
functions of the material parameters, such as chemical composition and
impurity concentration, in the neighborhood of the point $\r$.)

The results (\ref{eq:B203},\ref{eq:B204},\ref{eq:B20},\ref{eq:B21})
can be generalized to the case when $\phi$ and $\psi$ vary in space.
Equation (\ref{eq:B201}) is still valid in this case. If the length
scale of the variation is sufficiently large, the entire system can be
considered, within the standard approach, as consisting of small
subsystems weakly interacting with each other at imaginary borders.
Each of these subsystems, if allowed, will reach a local equilibrium
state which depends on the local values $n(\r)$ and $\epsilon(\r)$. We
assume, of course, that the magnetic field is constant. One can then
introduce local thermodynamic parameters (chemical potential,
temperature, entropy density, etc.) which are functions of the two
independent variables, $n$ and $\epsilon$. The values of extensive
thermodynamic parameters will be then given by the integral of the
corresponding local densities over the sample area.  In particular,
for the total energy, total entropy and total number of particles we
have
\begin{eqnarray}
E_T  \equiv  \langle H_T\rangle & = & \int[\epsilon\rp+
n\rp\phi\rp+
\epsilon\rp\psi\rp]\;d^3r,
\label{eq:34}
\\
S & = & \int s\left(\epsilon\rp,n\rp\right)\;d^3r,
\label{eq:28}
\\
N & = & \int n\rp\; d^3r,
\label{eq:29}
\end{eqnarray}
where $s(\epsilon, n)$, the local entropy density, is the same
function of $\epsilon$ and $n$ as for a uniform system.  The accuracy
of this approximation is limited by terms $(\nab\epsilon)^2$, $(\nab
n)^2$ which reflect the interaction between adjacent subsystems.  

It is more convenient to choose, instead of $\epsilon$ and $n$, a
local ``internal'' temperature $T\rp$ and chemical potential $\mu\rp$
as independent thermodynamic variables. Following Luttinger, we define
these parameters in the same way as it is conventionally done for a
homogeneous system
\begin{equation}
T^{-1}  =  \left.{\partial s\over\partial\epsilon}
\right|_n,\qquad
\mu=-T \left.{\partial s\over\partial n}
\right|_\epsilon.
\label{eq:30}
\end{equation}
[Of course, these definitions are consistent with the thermodynamic
relation~(\ref{eq:B43}), interpreted in a local sense, with $V$ and
$B$ fixed.]

The equilibrium state defined by the density matrix (\ref{eq:B201}) is
the state which maximizes the functional $\Phi\equiv S-\beta(E_T-\xi
N)$. Using (\ref{eq:34}), (\ref{eq:28}) and (\ref{eq:29}), and setting
the variational derivatives of $\Phi$ equal to zero, we find
\begin{eqnarray}
 & & T^{-1}\rp   =  \beta [1+\psi\rp] , 
\label{eq:B207}
\\
 & & \mu\rp  = 
[\xi-\phi\rp]/ [1+\psi\rp] , 
\label{eq:B208}
\end{eqnarray}
which is the generalization of (\ref{eq:B203}) and (\ref{eq:B204}) to
the non-uniform case. The equilibrium conditions may be alternatively
written as
\begin{eqnarray}
& & \delta(\mu/T)  =  -\phi\rp/T_0, 
\label{eq:B209}
\\
& & \delta(1/T)  = 
\psi\rp/T_0, 
\label{eq:B210}
\end{eqnarray}
where $T_0^{-1}=\beta$ is the unperturbed temperature of the sample.
If $T_0$ is replaced by the perturbed temperature $T$, these
conditions reduce to the conditions for thermal equilibrium obtained
by Luttinger\cite{lut}, which are valid for small changes in $\phi$
and $\psi$.

We now consider the magnetization currents. By definition, the
transport currents vanish in an equilibrium state, so the
magnetization currents are the total local currents in that case.  If
$\phi(\r)$, $\psi(\r)$, $T\rp$ and $\mu\rp$ vary sufficiently slowly
in space, it is clear that $\M^N\rp$ and $\M^E\rp$ will be still given
by Eqs.~(\ref{eq:B20}) and~(\ref{eq:B21}), provided that the arguments
$\mu$ and $T$ of the functions $\M_0^N$ and $\M_0^E$ are evaluated at
the position $\r$. (Corrections will be at least second order in the
gradients, if one is away from the boundary of a sample.) Then using
Eqs.~(\ref{eq:B3}) and~(\ref{eq:B4}), for a sample with uniform
chemical composition, far from the boundaries, we find
\widetext
\top{-2.8cm}
\begin{eqnarray}
\J_{{\rm mag}}^{{\rm bulk}} & = &
-{\partial\M_0^N\over\partial\mu}\times\nab\mu -
{\partial \M_0^N\over\partial T}\times\nab T-\M_0^N\times\nab\psi,
\label{eq:B22}
\\
\J_{{\rm mag}}^{E,{\rm bulk}} & = &
-{\partial\M_0^E\over\partial\mu}\times\nab\mu-
{\partial \M_0^E\over\partial T}\times\nab T
-\M_0^N\times\nab\phi-2\M_0^E\times\nab\psi,
\label{eq:B23}
\end{eqnarray}
\bottom{-2.7cm}
\narrowtext
\noindent
to leading order in $\phi$ and $\psi$.  There will also be additional
magnetization currents at the boundaries, given by Eqs.~(\ref{eq:B5}),
(\ref{eq:B6}), (\ref{eq:B20}) and~(\ref{eq:B21}).

\subsection{Non-Equilibrium States}

We now consider the non-equilibrium situation, where Eqs.
(\ref{eq:B209}) and (\ref{eq:B210}) are not satisfied and where
$\phi$, $\psi$, $T$, and $\mu$ may be in general time-dependent. The
local temperatures $T\rp$ and $\mu\rp$ are defined to be the same
functions of $\epsilon\rp$ and $n\rp$ as in the equilibrium case. We
continue to define the magnetization currents and magnetizations by
(\ref{eq:B20}), (\ref{eq:B21}), (\ref{eq:B22}) and (\ref{eq:B23}), and
we define the remaining contributions to the currents to be the
transport contributions.

By the locality hypothesis, the transport currents must be given by a
sum of terms proportional to the gradients $\nab\phi$, $\nab\psi$,
$\nab T$ and $\nab\mu$, in the limit where the applied fields are
small.  Our aim is to determine the coefficients of these terms. The
requirement that the transport currents vanish when (\ref{eq:B209})
and (\ref{eq:B210}) are satisfied means that the gradients enter only
in the combinations $[\nab\phi+T\nab(\mu/T)]$ and
$[\nab\psi-T\nab(1/T)]$. This observation is the generalization of the
``Einstein relation'' to the case in which there is a non-uniform
$\psi$, and allows the response to the statistical fields $\mu$ and
$T$ to be related to the response to the mechanical fields $\phi$ and
$\psi$.

To determine the remaining unknown coefficients, we consider a
particular situation, where the potentials $\phi$ and $\psi$ vary
periodically in time, with a characteristic frequency $\omega$ small
compared to the microscopic relaxation rate $\tau_m^{-1}$ but large
compared to the size-dependent relaxation rates $\tau_M^{-1}$ for the
energy and particle density. (This is the ``rapid case'' of
Ref.~\onlinecite{lut}.) In this situation the local values of
$\epsilon$ and $n$ are not changed from their initial values, so that
$\mu$ and $T$ remain constant throughout the sample.

We concentrate on a region far from the boundaries. Since 
$\nabla T=\nabla\mu=0$, we may write the total 
currents as
\begin{eqnarray}
\J^{\bulk} & = & \hat{L}^{(1)}(-\nab\phi)+
\hat{L}^{(2)}(-\nab\psi),
\label{eq:11}
\\
\J^{E,\bulk} & = & \hat{L}^{(3)}(-\nab\phi)+
\hat{L}^{(4)}(-\nab\psi),
\label{eq:12}
\end{eqnarray}
with transport coefficients that can be expressed in terms of
time-dependent correlation functions in the framework of the standard
Kubo formula\cite{lut}:
\begin{eqnarray}
L_{\alpha\gamma}^{(1)} & = &
\lim_{s\rightarrow 0}
{1\over V}\int_0^\infty \!\!\!\!\!\!
dt\; e^{-st}\int_0^\beta \!\!\!\!\!
d\beta'\langle j_0^\gamma(-t-i\beta')
j_0^\alpha(0)\rangle,
\label{eq:13}
\\
L_{\alpha\gamma}^{(2)} & = & 
\lim_{s\rightarrow 0}
{1\over V}\int_0^\infty \!\!\!\!\!\!
dt\; e^{-st}\int_0^\beta \!\!\!\!\!
d\beta'\langle j_0^{\gamma E}(-t-i\beta')
j_0^\alpha(0)\rangle,
\label{eq:14}
\\
L_{\alpha\gamma}^{(3)} & = & 
\lim_{s\rightarrow 0}
{1\over V}\int_0^\infty\!\!\!\!\!\!
dt \; e^{-st}\int_0^\beta\!\!\!\!\!
d\beta'\langle j_0^\gamma(-t-i\beta')
j_0^{\alpha E}(0)\rangle,
\label{eq:15}
\\
L_{\alpha\gamma}^{(4)} & = &
\lim_{s\rightarrow 0}
{1\over V}\int_0^\infty \!\!\!\!\!\!
dt\; 
e^{-st}\int_0^\beta \!\!\!\!\!
d\beta'\langle
j_0^{\gamma E}(-t-i\beta') j_0^{\alpha E}(0)\rangle.
\label{eq:16}
\end{eqnarray}
Here the subscript 0 on each time-dependent current operator $\j_0$
and $\j_0^E$ indicates the $q\rightarrow 0$ limit of its spatial
Fourier transform, and the angular brackets denote the quantum
mechanical and thermodynamic average in the equilibrium state of an
infinite system at temperature $\beta^{-1}=T$ and chemical potential
$\mu$.

In order to find the {\it transport currents} when $\nabla T=\nabla
\mu=0$, we subtract from (\ref{eq:11}) and (\ref{eq:12}) the bulk
magnetization currents, given by (\ref{eq:B22}) and (\ref{eq:B23}). We
thus find
\begin{eqnarray}
\J_{{\rm tr}} & = &
-\hat L_1^{(1)}\nab\phi-
[\hat L^{(2)}-\M_0^N\cdot\hep]\nab\psi , 
\label{eq:B24}
\\
\J_{{\rm tr}}^E & = &
-(\hat L^{(3)}-\M_0^N\cdot\hep)\nab\phi- [\hat
L^{(4)}-2\M_0^E\cdot\hep]\nab\psi ,
\label{eq:B25}
\end{eqnarray}
where $\hep$ is the unit antisymmetric three-tensor:
$(\hat\epsilon)_{\alpha\beta\gamma}= 1(-1)$ if the $\alpha\beta\gamma$
is an even (odd) permutation of $xyz$, and zero if two indices are
equal.  By definition, transport currents flow only in the bulk of the
sample.  To simplify notation, we have therefore dropped the label
``bulk'' in the above equations, and will do so for all subsequent
transport current densities.

Now we can consider the general case, for an arbitrary value of
$\omega\tau_M$, where $\nab\mu$, $\nab T$, $\nab\psi$, and $\nab\phi$
may be all non-zero.  As remarked earlier, the condition that the
transport currents vanish in equilibrium requires that the gradients
enter only in the combinations $[\nab\phi+T\nab(\mu/T)]$ and
$[\nab\psi-T\nab(1/T)]$. (In other words, the Einstein relations,
extended to the case where $\nab T$ and $\nab\psi$ are different from
zero, apply to the {\it transport} currents in the presence of finite
magnetic field.) Thus we have, in the general case:
\widetext
\top{-2.8cm}
\begin{eqnarray}
\J_{{\rm tr}} & = & 
-\hat L^{(1)}[\nab\phi+T\nab(\mu/T)]
+(\hat L^{(2)}-\M_0^N\cdot\hep)[-\nab\psi+ T\nab(1/T)] ,
\label{eq:B26}
\\
\J_{{\rm tr}}^E & = & 
-(\hat L^{(3)}-\M_0^N\cdot\hep) [\nab\phi+T\nab(\mu/T)] +(\hat
L^{(4)}-2\M_0^E\cdot\hep)[-\nab\psi+ T\nab(1/T)].
\label{eq:B27}
\end{eqnarray}
\bottom{-2.7cm}
\narrowtext
\noindent
Of course, (\ref{eq:B26}) and (\ref{eq:B27}) reduce to the Luttinger
formulae\cite{lut} when the magnetic field is absent, so that
$\M_0^N=\M_0^E=0$.

To obtain the local currents, in a uniform sample far from the
boundaries, in the general case where $\nab T$, $\nab\mu$, $\nab\phi$
and $\nab\psi$ are all independent, we must add the magnetization
currents, given by (\ref{eq:B22}) and (\ref{eq:B23}) to the transport
currents given by (\ref{eq:B26}) and (\ref{eq:B27}). In addition,
there will be boundary currents from the magnetization. In the absence
of the fictitious gravitational potential $(\psi=0)$ the change in the
integrated boundary currents at a point $\bbox{\tau}$,
Eqs.~(\ref{eq:B5})~(\ref{eq:B6}), is determined by the changes in
the temperature and chemical potential at that point, according to
\begin{eqnarray}
\delta\I & = &
-\hn\times\left[{\partial\M_0^N\over\partial\mu}\delta\mu+
{\partial\M_0^N\over\partial T}\delta T\right] ,
\label{eq:B28}
\\
\delta \I^E & = & 
-\hn\times\left[{\partial\M_0^E\over\partial\mu}\delta\mu+
{\partial \M_0^E\over\partial T}\delta T+
\phi\rp \M_0^N\right] .
\label{eq:B29}
\end{eqnarray}
Although we have introduced the fictitious field $\psi$ to derive
linear response, we will have no further need of it, and hence {\it
shall set $\psi=0$ for the remainder of this paper}.

It is convenient to introduce the electrochemical potential, defined
by
\begin{equation}
\xi\rp\equiv
\mu\rp+\phi\rp.
\label{eq:B30}
\end{equation}
An ideal voltmeter, (with leads that have no thermopower) will measure
the difference in $\xi$ between two contact points. In a thermodynamic
equilibrium state, the value of $\xi$, as well as the temperature $T$,
will be constant throughout the system.

It is also convenient to define a transport heat current density as
\begin{equation}
\J_{{\rm tr}}^Q\equiv
\J_{{\rm tr}}^E-\xi
\J_{{\rm tr}}.
\label{eq:B31}
\end{equation}
Then we may rewrite (\ref{eq:B26}) and (\ref{eq:B27}) in the form
\begin{eqnarray}
\J_{{\rm tr}} & = & -\hat N^{(1)}\nab\xi-\hat N^{(2)}
(\nab T)/T,
\label{eq:B32}
\\
\J_{{\rm tr}}^Q & = & -\hat N^{(3)}\nab\xi-\hat N^{(4)}
(\nab T)/T,
\label{eq:B33}
\end{eqnarray}
where
\widetext
\top{-2.8cm}
\begin{eqnarray}
\hat N^{(1)} & = & \hat L^{(1)}.
\label{eq:B34}
\\
\hat N^{(2)} & = & \hat L^{(2)}
-\mu\hat L^{(1)}-\M_0^N
\cdot\hep,
\label{eq:B35}
\\
\hat N^{(3)} & = & \hat L^{(3)}
-\mu\hat L^{(1)}-\M_0^N
\cdot\hep,
\label{eq:B36}
\\
\hat N^{(4)} & = & \hat L^{(4)}
-\mu(\hat L^{(3)}+\hat L^{(2)})+
\mu^2\hat L^{(1)}-2(\M_0^E-
\mu\M_0^N)\cdot\hep.
\label{eq:B37}
\end{eqnarray}
\bottom{-2.7cm}
\narrowtext
\noindent
Although our derivation has assumed $\phi$ to be infinitesimal, the
final results for the transport currents, given by
(\ref{eq:B30})--(\ref{eq:B37}) and (\ref{eq:13})--(\ref{eq:16}) are
written in a form that remains valid when $\phi$ is not small.  This
accounts for the slight differences (which are beyond leading order in
the driving fields) between the above coefficients, $\hat N^{(i)}$,
and those obtained by direct substitution of Eqs.~(\ref{eq:B26})
and~(\ref{eq:B27}) into Eq.~(\ref{eq:B31}). When $\phi$ is not small,
the expressions (\ref{eq:11}) and~(\ref{eq:12}) should contain
additional higher-order terms in $\phi$; the coefficients $\hat L$
appearing above cannot therefore be defined by these expressions, but
are assumed to be defined by the Kubo formulae
(\ref{eq:13},\ref{eq:14},\ref{eq:15},\ref{eq:16}).  In the case of
finite $\phi$, equation~(\ref{eq:B23}) for the bulk magnetization
current $\J_{{\rm mag}}^{E,\bulk}$ must also be modified by the
addition to the right-hand side of a term $\phi\J_{\rm mag}^{\bulk}$;
the edge current $\I^E$ is still given by (\ref{eq:B6}) and
(\ref{eq:B21}).

\subsection{Onsager Relations}
\label{subsec:onsager}

The transport coefficients for $\J_{{\rm tr}}$ and $\J_{\rm tr}^Q$,
given by (\ref{eq:B32})--(\ref{eq:B37}), obey Onsager symmetry
relations\cite{onsager1,onsager2} of the form
\begin{eqnarray}
N_{\alpha\gamma}^{(1)}(\B) & = &
N_{\gamma\alpha}^{(1)}(-\B),
\label{eq:B38}
\\
N_{\alpha\gamma}^{(2)}(\B) & = &
N_{\gamma\alpha}^{(3)}(-\B)
\label{eq:B39},
\\
N_{\alpha\gamma}^{(4)}(\B) & = &
N_{\gamma\alpha}^{(4)}(-\B).
\label{eq:B40}
\end{eqnarray}
To see that this is the case, we first establish that the coefficients
$L_{\alpha\gamma}^{(i)}$ $(i=1,2,3,4)$ obey Onsager relations of the
same form as (\ref{eq:B38})--(\ref{eq:B40}).  This follows from the
expressions relating $L_{\alpha\gamma}^{(i)}$ to the current
correlation functions, (\ref{eq:13})--(\ref{eq:16}), and the
invariance of the Hamiltonian $H_0$ under simultaneous reversal of
time and magnetic field. Secondly, we note that the magnetizations
$\M^N$ and $\M^E$ reverse sign under reversal of $\B$. Equations
(\ref{eq:B38})--(\ref{eq:B40}) follow directly.

In contrast the {\it local} current densities do not satisfy the
Onsager relations.  The local current densities differ from the local
transport current densities by the bulk ``magnetization currents''.
These additional magnetization contributions to the local response in
general depend on which driving field is applied (e.g.  whether it is
$\nabla\mu$ or $\nabla\phi$), and give rise to a set of transport
coefficients, one for each driving field, which differ by terms
proportional to the gradients of magnetization densities.  Since no
general symmetries relate these additional magnetization terms,
neither the Einstein relations nor the Onsager relations hold locally
for the total current response.

As a specific example, consider the local bulk currents $\J$ and
$\J^Q\equiv\J^E-\xi\J$ in terms of $\nab\mu$ and $\nab T$, under
conditions where $\nabla\phi=\nabla\psi=0$. By combining
(\ref{eq:B31})--(\ref{eq:B37}), with (\ref{eq:B1}), (\ref{eq:B2}),
(\ref{eq:B22}) and (\ref{eq:B23}), one may readily obtain expressions
for the appropriate coefficients ${\cal L}_{\alpha\gamma}^{(i)}$
\begin{eqnarray}
\hat{\cal L}^{(2)} & = & \hat N^{(2)}+T
{\partial\M_0^N\over\partial T}
\cdot\hep ,
\label{eq:B41}
\\
\hat{\cal L}^{(3)} & = & \hat N^{(3)}+
\left(
{\partial\M_0^E\over\partial \mu}
-\mu{\partial\M_0^N\over\partial\mu}\right)
\cdot\hep .
\label{eq:B42}
\end{eqnarray}
As far as we are aware, there is no symmetry relating the derivatives
of $\M_0^E$ and $\M_0^N$ with respect to $\mu$ and $T$, so these
coefficients apparently do not satisfy the usual Onsager relations.

Moreover, if we consider instead a situation where $\nab\mu=\nab T=0$,
but $\nab\phi\not=0$, we find according to (\ref{eq:11}) and
(\ref{eq:12}), $\J^Q=-(\hat L^{(3)}-\mu\hat L^{(1)})\nab\phi$.  This
coefficient, which is different from $\hat{\cal L}^{(3)}$, is also
clearly not related to $\hat{\cal L}^{(2)}$ by Onsager symmetry in the
general case.

Although the contribution of the magnetization current does not appear
in standard transport experiments, the local electric current
distribution is measurable, at least in principle, by a sufficiently
sensitive measurement of the magnetic field generated by currents in
the sample.  Since the magnetization current does not dissipate Joule
heat, it could not be detected using the well-known technique based on
local luminescence intensity\cite{klass}.  Measurements of the local
electric field distribution by means of the electro-optic
effect\cite{fontein} do not serve this purpose either, since the
magnetization current is related to the ``statistical fields''
$\nabla\mu$ and $\nabla T$ rather than to the electric field
$(1/e)\nabla\phi$.  In the case of the energy current or heat current,
we are not aware of any reasonable method for direct measurement of
the local currents.

We illustrate our results further with the example of the number
current $\J$ for a two-dimensional electron system with no disorder,
at a Landau-level filling fraction $\nu^*$ for which the electron
system is incompressible at $T=0$. At the filling fraction $\nu^*$, we
have $n=\nu^*e|B|/hc$. Then we have
\begin{equation}
\left.{\partial\M^N\over\partial\mu}
\right|_{T,B}=-{c\over e}
\left.{\partial n\over\partial\B}\right|_{T,\mu}
\rightarrow -{\nu^*\over h}
\hz\;{\rm sgn}(B_z),
\label{eq:B47}
\end{equation}
where the last equality follows from the incompressibility condition
for $T\rightarrow 0$. Here $\hz$ is a unit vector directed upward from
the plane.

The coefficient $\hat L^{(1)}$ for this system is given by the Hall
conductivity: $\hat L^{(1)}=\left({\nu^*\over h}\right) {\rm
sgn}(B_z)(\hz\cdot\hep)$. On the other hand, substituting
$\xi=\mu+\phi$ into Eqs. (\ref{eq:B1}), (\ref{eq:B22}) and
(\ref{eq:B26}), we see that the coefficient describing the local
response to $\nab\mu$ vanishes, and all number current flow in the
bulk is due to the electric field: $\J^{\bulk}=-(\nu^*/h){\rm
sgn}(B_z)[\hz\times\nab\phi]$. The current driven by an inhomogeneous
chemical potential is localized at the edge, and is given by
$\delta\I=(\nu^*/h){\rm sgn}(B_z)\delta\mu\rp[\hn\times\hz]$, where
$\hn$ is a unit vector in the plane, perpendicular to edge, in the
outward direction.

\subsection{Inhomogeneous Samples}

Although we have concentrated so far on the case of a homogeneous
sample with boundaries that are sharp compared with the overall length
scale, it is easy to generalize our results to the case of a sample
whose material parameters, such as chemical composition, vary on a
macroscopic length scale. The formulae~(\ref{eq:B32})
and~(\ref{eq:B33}) still hold locally for the transport currents in
this case, with the qualification that the transport coefficients
$\hat N^{(i)}$ depend on the local material parameters, and can
therefore vary from one place to another in the sample. The
magnetization currents at the boundary of the sample are still given
by (\ref{eq:B28}) and (\ref{eq:B29}), with the qualification that the
functions $\M_0^N$ and $\M_0^E$ may also vary from one place to
another because of their implicit dependence on the local material
parameters.  The magnetization currents in the bulk of the sample are
no longer given simply by Eqs.~(\ref{eq:B22}) and~(\ref{eq:B23})
however. If we denote the material parameters by a set of variables
$\{\eta_i\}$, and we set $\nabla\psi=0$, then (\ref{eq:B22}) and
(\ref{eq:B23}) should be replaced by
\widetext
\top{-2.8cm}
\begin{eqnarray}
\J_{\rm mag}^{\bulk} & = & 
- {\partial \M_0^N\over\partial\mu}\times\nab\mu-
{\partial\M_0^N\over\partial T}\times\nab T
-\sum_i{\partial\M_0^N\over\partial\eta_i}\times\nab\eta_i ,
\label{eq:B211}
\\
\J_{\rm mag}^{E,\bulk} & = & 
-{\partial \M_0^E\over\partial\mu}\times\nab\mu-
{\partial\M_0^E\over\partial T}\times\nab T
-\sum_i{\partial\M_0^E\over\partial\eta_i}\times\nab
\eta_i-\M_0^N\times\nab\phi+\phi\J_{\rm mag}^{\bulk} .
\label{eq:B212}
\end{eqnarray}
\bottom{-2.7cm}
\narrowtext
\noindent
[These expressions follow directly from (\ref{eq:B3}), (\ref{eq:B4}),
(\ref{eq:B20}) and (\ref{eq:B21}).]  In the quantities
$\partial\M_0^N/\partial\eta_i$ and $\partial\M_0^E/\partial\eta_i$,
which appear in (\ref{eq:B211}) and (\ref{eq:B212}), it is necessary
to keep not only the zeroth order terms, but also the first-order
changes engendered by the deviations $\delta\mu\rp$ and $\delta T\rp$.

\subsection{Long-Range Forces}

The derivation in the previous subsections, and various intermediate
results, require modification when there are long-range forces due to
unscreened Coulomb interactions.  In two-dimensional structures, this
is the case when the metallic gate is either absent or is situated
farther than the characteristic length scale of the fields applied.

If uncompensated electric charges are present, there can be energy
transport over long distances via the macroscopic electric field, and
the total energy current at a given point of space does not, in
general, depend solely on the state of the particle system in the
immediate neighborhood of that point.  Moreover, the convention of
Eq.~(\ref{eq:B19}), in which the interaction contribution to the
energy current is concentrated along the line-segments joining each
pair of particles, is not generally used in this case.

The most convenient approach is to break up the interparticle
interactions into a long-range piece, mediated by the macroscopic
electric field $\E\rp$, and a short-range piece, which includes
everything else.  The macroscopic field is supposed to be averaged
over a region sufficiently large that fluctuations in the field,
arising from thermal or quantum fluctuations in the microscopic
electronic charge density, can be neglected.  Thus there is no entropy
transport via the macroscopic field.  As a result, we find that with
appropriate definitions we can write local hydrodynamic equations for
the {\it heat} current and particle current which are similar in form
to the equations derived for short-range forces. (We assume here that
we are working at a temperature sufficiently low that heat transport
via the radiation field may be completely neglected.  Formally, this
assumption is imposed by taking the limit where the speed of light $c$
is infinite.)

To make these points clearer, we note that since charge is locally
conserved, charge fluctuations with the longest range effects are
electric dipole fluctuations.  In the absence of the radiation field,
the interaction between dipoles at two different points falls off as
the inverse cube of the separation, and the rate of energy transfer
due to random thermal motion would be expected to fall off as the
inverse sixth power of the separation.  At large distances this
process will be much slower than the conventional process of heat
conduction (already taken into account in our discussions) wherein
energy is transported diffusively via a series of many short jumps.

To proceed formally, we redefine the interaction $u_{ij}$ appearing in
subsection~B to include only the short-range part of the Coulomb
interaction, after effects of the macroscopic electric field $\E$ are
subtracted out.  First, we choose a truncation radius $r_{\rm cut}$
which is much larger than the average inter-particle distance but much
smaller than the macroscopic scales of external fields and, if the
sample is not uniform, the scale of the equilibrium density variation.
We split the Coulomb interaction potential into the sum of two terms,
$u_{ij}^{\rm tot}({\r})=u_{ij}^{{\rm short}}(\r)+ u_{ij}^{\rm
long}(\r)$, where $u_{ij}^{\rm short}({\r})$ decays rapidly at $r\gg
r_{\rm cut}$, and $u_{ij}^{\rm long}(\r)$ contains the long-range tail
of the interaction and changes smoothly at distances $r\ll r_{\rm
cut}$.  Then we replace $u_{ij}$ by $u_{ij}^{\rm short}$ in the
definition of $H_0$ and $h_i$, Eq.~(\ref{eq:1}), and include the
long-range component $u_{ij}^{\rm long}({\r})$ in the self-consistent
macroscopic field ${\bbox E}$.

We continue to define $h\rp$ by Eq.~(\ref{eq:3}), and define the
internal energy density $\epsilon(\r)$ as the expectation value
$\langle h\rp\rangle$. Then $\epsilon$ may be interpreted as the
matter contribution to the energy density. The total energy (with
$\psi=\phi=0$) is then given by
\begin{equation}
E=\int d^3\r\left[
\epsilon+{\kappa|\E|^2\over 8\pi}+{|\B|^2\over 8\pi}\right] ,
\label{eq:B48}
\end{equation}
where $\kappa$ is the dielectric constant of the background material,
and we assume the background magnetic permeability is unity. The
energy current will similarly be broken up into two parts
\begin{equation}
\j_{{\rm tot}}^E=\j^E+{c\E\times\B\over 4\pi} ,
\label{eq:B49}
\end{equation}
where the second term is the standard contribution from the
macroscopic electromagnetic fields, and $\j^E$, which we may think of
as the matter contribution to the energy current, is defined by
Eqs.~(\ref{eq:B19})--(\ref{eq:21}), with $u_{ij}$ replaced by
$u_{ij}^{\rm short}$ in the definition of $F_{ij}^\alpha$.  Of course,
$\E$ and $\B$ are determined self-consistently from the macroscopic
current and charge distributions using Maxwell's equations in the
static limit.  Equations~(\ref{eq:B48}) and~(\ref{eq:B49}) are
asymptotically correct in the limit where $r_{\rm cut}$ is large
compared to the microscopic scale but small compared to the scale of
variation of $\E$.

For a two-dimensional electron system in an external magnetic field,
the magnetic fields arising from currents in the sample are generally
very small, and may be omitted from the term $|\bbox{B}|^2/8\pi$ in
Eq.~(\ref{eq:B48}).  Thus this term is independent of the state of the
electron system and may be ignored if desired. On the other hand, the
magnetic fields generated by the currents in the sample must be
included in the second term on the right-hand side of (\ref{eq:B49}),
because the speed of light appears as a prefactor. Note also that
$\epsilon$ and $\j^E$ are restricted to the two-dimensional layer, but
the electromagnetic contributions to~(\ref{eq:B48}) and~(\ref{eq:B49})
extend into the space outside.  Below we focus on the matter part
${\bbox j}^E$ of the energy current.

It is now possible to redo the arguments of the previous section with
little modification. We restrict our attention to the situation where
the applied magnetic field is independent of time and assume that the
macroscopic electric field may be derived from a scalar potential
\begin{equation}
\E(\r)=-\nab\Phi\rp.
\label{eq:B50}
\end{equation}
The potential $\Phi$ is obtained self-consistently, and includes
effects of any macroscopic time-dependent variations in $n(\r)$, as
well as the effects of any static charges present in equilibrium and
the external perturbations embodied in $\phi$.  In general, we cannot
consider that $\Phi$ is infinitesimal, even in equilibrium.

In order to repeat our previous derivations, one needs to consider a
non-zero gravitational potential $\psi$; we present here the final
result, and hence set $\psi =0$.  We redefine the electrochemical
potential as
\begin{equation}
\xi\rp=\mu\rp-e\Phi\rp.
\label{eq:B52}
\end{equation}
The definition (\ref{eq:B18}) of the energy current in the perturbed
system is now replaced by
\begin{equation}
\label{eq:B53}
\J^E=\j^E-e\Phi\J ,
\end{equation}
where $\J=\j$ is the particle current density, and $\j^E$ was defined
above. As we did in previous sections, we split the local currents
$\J$ and $\J^E$ into magnetization and transport parts, and we define
a transport heat current as given by Eq.~(\ref{eq:B31}).  Then
$\J_{\rm tr}^Q$ and $\jtr$ obey equations identical
to~(\ref{eq:B32})--(\ref{eq:B37}), with $L_{\alpha\gamma}^{(i)}$
defined by Eqs.~(\ref{eq:13})--(\ref{eq:16}) in terms of the
correlators for the matter currents $\j(\r)$ and $\j^E(\r)$ under the
Hamiltonian $H_0$ for a uniform system in equilibrium, with
$\phi=\psi=0$.  The transport coefficients $N_{\alpha\gamma}^{(i)}$
obey the same Onsager relations as before.

Note that different definitions of the macroscopic electric field, as
may be obtained by different choices of the truncation radius $r_{\rm
cut}$, will generally cause an exchange of contributions between the
chemical potential $\mu(\r)$ and the electrostatic potential
$\Phi(\r)$. This will also transfer contributions between the first
and second terms in Eq.~(\ref{eq:B53}), leaving the sum $\J^E$
unchanged.  As long as the different values of $r_{\rm cut}$ remain
sufficiently large, the change in the truncation should not affect the
coefficients $\hat L^{(i)}$ since the correlators in
Eqs.~(\ref{eq:13})--(\ref{eq:16}) are sensitive only to short-range
properties of the system.

We also note that in realistic two-dimensional systems, the component
of the electric field perpendicular to the electron layer, arising
from charges on gates or from ionized impurities displaced from the
layer, may play an important role in confining the electrons to the
layer.  It is therefore important to include this part of the
macroscopic electric field in the unperturbed Hamiltonian $H_0$ and in
the definition of the energy current $\j^E$ when calculating the
correlation functions that appear in~(\ref{eq:13})--(\ref{eq:16}).
Formally this can be done by including the perpendicular confining
field in the one-body potential $V$ which enters Eq.~(\ref{eq:1}), and
excluding it from the macroscopic potential $\Phi(\r)$.  Then the
equilibrium $\Phi(\r)$ is a constant in the direction perpendicular as
well as parallel to the layer, and can be safely omitted from $H_0$.
In fact, it may be convenient to include the entire equilibrium value
of the electrostatic potential in $V$, even in an inhomogeneous
system, so that $\Phi$ describes only long-wavelength fluctuations
about equilibrium.

\subsection{Additional Remarks}

Throughout this section, we have treated the fields $\mu$, $T$, $\psi$
and $\phi$ (or $\Phi$, in subsection~G) as independent variables,
which can be arbitrary functions of position, subject to the
constraints that gradients are small, and that $\phi$ and $\psi$ are
infinitesimal. (The fictitious field $\psi$ was set equal to zero in
the latter part of the section.)  The transport currents depend on
gradients of $\mu$ and $\phi$ (or $\Phi$) only through the combination
$\nab\xi$.  In a d.c. transport experiment, there are strong
additional constraints arising from the requirements that the currents
must be divergence-free in the interior of the sample, and must
satisfy appropriate constraints at the boundaries.  Typically, these
conditions completely determine the spatial variations of $\xi$ and
$T$ throughout the sample interior, when boundary values of the fields
are specified, or when current flows through the boundaries are given.

For the case of a two-dimensional electron system on a
three-dimensional substrate, the situation is slightly more
complicated. We shall be concerned with situations where the substrate
is an electrical insulator, so that the divergence of the
two-dimensional electric current is required to be zero in the
analysis of experiments. On the other hand, we consider the thermal
coupling to the substrate, via absorption or emission of phonons, to
be small but not zero. Then on length scales large compared to
$(D_T\tau_{ep})^{1/2}$, where $\tau_{ep}$ is the electron-phonon
relaxation time and $D_T$ an appropriate thermal diffusion constant
for the isolated electron system, the divergence of $\J^Q$ is not
necessarily zero. Instead, one should take the value of $T\rp$ to be
an independent variable determined by conditions in the substrate.

Although our previous discussions assumed the sample to be isolated
from its environment except at its edges, the transport equations
derived above should remain valid provided the electron-phonon
coupling is sufficiently weak that $\tau_{ep}$ is large compared to
the microscopic times necessary to establish local equilibrium in the
electron system.

In order to calculate the transport currents, it is not generally
necessary to find the separate portions of $\nab\xi$ arising from
$\nab\mu$ and from the electric field. This is necessary, however, if
one wishes to obtain the local current distribution.  In practical
situations, where the nearest external conductor is far from the
electron layer compared to the mean spacing between electrons in the
layer, the value of $|\nab\mu|$ will be relatively small compared to
the value of $|\nab\phi|$ or $|\nab\Phi/e|$.  This is due to the fact
that, in the absence of external screening, perturbations in $\phi$
and $\mu$ are not really independent and their characteristic
magnitudes can be expressed via each other. Suppose that a
non-equilibrium perturbation in the chemical potential $\delta\mu(\r)$
with a large length scale $l_{\mu}$ is created in a two-dimensional
system.  The resulting variation in the particle density has a
magnitude $\delta n \sim \delta\mu /(d\mu /dn)$.  The magnitude of the
potential variation caused by this accumulation and depletion of
electrons can be estimated as $\delta\phi \sim (e^2/\kappa)\delta n
l_{\mu}$.  Thus, in a compressible system, we find the ratio of
gradients $\nabla\mu /\nabla\phi$ to be of the order of $\sim \delta
\mu /\delta\phi \sim r_s/l_{\mu}$ which is small in the limit of large
$l_{\mu}$.  In an incompressible system (for instance, a
macroscopically wide strip of a system in the middle of a quantized
Hall plateau, if there are no localized states in the energy gap), the
accumulation of electrons due to the perturbation in the chemical
potential occurs only at the edges of the system. The accumulated
charge creates an electric field that slowly (as $1/r$) vanishes into
the interior of the sample. The resulting ratio $\delta\mu/\delta\phi$
is as small as $1/\ln(W/r_s)$ where $W$ is the width of the
strip\cite{wexlerthouless}.

The above observation has an impact on the issue of the
edge-versus-bulk current distribution. Since, in the absence of a
temperature gradient, the boundary currents arise solely due to
$\delta\mu$, and the bulk currents are due to both $\delta\mu$ and
$\delta\phi$, the non-equilibrium current in a system without gates
flows predominately in the bulk. The situation changes, however, if
the temperature is not uniform.  As we shall see in Sec.~IV, the
boundary fraction of the net thermocurrent is at least as significant
as its bulk counterpart.

\section{LINEAR RESPONSE IN THE ABSENCE OF DISORDER}

In the previous section, the local current response to the electric
and statistical fields was expressed in terms of the coefficients of
the mechanical response in the bulk $\hat L^{(i)}$ defined by the
general expressions~(\ref{eq:13})--(\ref{eq:16}). Now we derive
explicitly three of the four response coefficients $\hat L^{(i)}$ for
a simple case of a uniform disorder-free sample. Since the internal
magnetization currents and the boundary currents are already expressed
in terms of the equilibrium magnetizations and their derivatives as
given by Eqs.~(\ref{eq:B45},\ref{eq:B46},\ref{eq:B52},\ref{eq:B53}),
here we consider only the ``transport'' components of the bulk
currents. We restrict our attention to the case of a two-dimensional
electron system, with magnetic field $B$ in the $\hz$ direction,
perpendicular to the layer.

Our main result is the expression for the number current density
\begin{equation}
\jtr={nc\over eB}\hat\epsilon\nab\xi
+{sc\over eB}\hat\epsilon\nab T,
\label{eq:69}
\end{equation}
which shows that, in the absence of disorder, the transport
contribution to the number current density is fully determined by the
equilibrium number and entropy densities. (In this section, the symbol
$\hat \epsilon$ represents the two-dimensional antisymmetric tensor
$\epsilon_{xy}=-\epsilon_{yx}=1$.)

Equation (\ref{eq:69}) is valid locally for any interacting electron
system, provided it does not have a shear modulus, i.e., is a fluid,
and the energy spectrum of electrons is quadratic. We present two
different ways to obtain this result.  In Sec.~A, we derive
Eq.~(\ref{eq:69}), and hence the coefficients $\hat L^{(1)}$ and $\hat
L^{(2)}$ directly, using arguments based on fluid dynamics. In the
alternative proof given in Sec.~B, we first derive the coefficients
$\hat L^{(1)}$ and $\hat L^{(3)}$ by studying the current response in
a uniform electric field, and then obtain $\hat L^{(2)}$ from the
Onsager symmetry. The advantage of the second derivation is that it
deals with a homogeneous non-equilibrium system, $\nabla
T=\nabla\mu=0$.

Both derivations that follow employ the physical notion of internal
pressure, which is not quite obvious in the presence of a magnetic
field and which we now discuss briefly. In the presence of a uniform
magnetic field $\B||\hz$ the equation for conservation of momentum on
the microscopic scale may be written
\begin{equation}
m{\partial\J\rp\over\partial t}
=-\rho\rp\nab V\rp -
{e\over c} \J\rp\times\B-
\nab\cdot\hat\pi\rp,
\label{eq:B101}
\end{equation}
where $V\rp$ is the one-body potential and $\hat \pi\rp$ is the
internal stress tensor at point $\r$. As in the case of the energy
current discussed in Section~II, there is not a unique definition of
$\hat\pi$ for a system with finite range forces. However, a definition
consistent with Eq.~(\ref{eq:B101}) and with the requirement of
quasi-locality is
\begin{equation}
\pi_{\gamma\alpha}\rp=
{m\over 4}\sum_i
\{v_i^\alpha,\{v_i^\gamma,\delta_i\}\}+
\frac{1}{2}\sum_{i\not=j}\tau_{ij}^{\gamma\alpha}\rp ,
\label{eq:B102}
\end{equation}
where $\tau_{ij}^{\gamma\alpha}$ is defined by Eq.~(\ref{eq:tau}).
Equation~(\ref{eq:B102}) may be checked by integrating both sides of
(\ref{eq:B101}) over an infinitesimal volume, to obtain the rate of
change of the momentum inside the volume.  The integral of the last
term on the right hand side of (\ref{eq:B101}) is equal to the
integral of the stress tensor over the surface enclosing the volume.
The first term of (\ref{eq:B102}) then gives the change in momentum
due to particles crossing the surface, while the second term gives the
force exerted on particles inside the volume by particles outside the
volume [cf. Eq.~(\ref{eq:chi})].

In applying the above equations to the present problem, we interpret
$m$ as the band mass rather than the bare mass of the electron, and
$V\rp$ excludes the periodic potential of the ions. Thus, when there
are no impurities present and no applied electric field, $V\rp$ is a
constant in the interior of the sample, and $\nab V$ arises only
from the confining potential at the boundaries. We also assume that
$u_{ij}$ depends only on the distance between the electrons, so that
$\F_{ij}$ is parallel to $(\r_i-\r_j)$ and the stress tensor is
symmetric. We specialize to the case of a two-dimensional
electron system and thus use notation appropriate to two dimensions.

In thermal equilibrium, far from the boundaries, in an isotropic
system, the stress tensor must be proportional to the unit tensor, so
we may write
\begin{equation}
\langle\pi_{\alpha\gamma}\rp\rangle=
P_{{\rm int}}\delta_{\alpha\gamma} ,
\label{eq:B103}
\end{equation}
where $P_{{\rm int}}$, the ``internal pressure'', depends on the
chemical potential and temperature.  In the presence of a magnetic
field, $P_{{\rm int}}$ is not equal to the pressure $P$ which appeared
in the thermodynamic equations (\ref{eq:B43})--(\ref{eq:B46}) and
which is equal to the force per unit length exerted by the boundaries
on the contained electron gas. Rather, we have
\begin{equation}
P_{{\rm int}}=P-MB .
\label{eq:70}
\end{equation}
The difference between $P_{{\rm int}}$ and $P$ arises from the Lorentz
force exerted by $B$ on the boundary current $\I=(c/e)\hn\times\M$.

Equation (\ref{eq:70}) may be obtained directly from equations
(\ref{eq:B101}) and (\ref{eq:B103}) if we integrate the right-hand
side of (\ref{eq:B101}) along a line-segment from a point $\r_1$ in
the interior of the sample, to a point $\r_2$ where $\rho\rp=0$,
passing through a point $\bbox{\tau}$ on the boundary. Since $\partial
\J/\partial t=0$ in equilibrium, and since $\hat\pi(\r)=0$ at point
$\r_2$, we find
\begin{equation}
P_{{\rm int}}+MB=
\int_{r_1}^{r_2}
\!\! d\r\cdot\rho\rp
\nab V\rp.
\label{eq:B104}
\end{equation}
The right-hand side of (\ref{eq:B104}) is just the force per unit
length exerted by the boundary at point $\bbox{\tau}$.
The fact that the pressure entering the thermodynamic equations is
indeed the same as the force per unit length exerted by the boundaries
follows from the well-known fact that a magnetic field constant in
time cannot produce work on charged particles. Hence, although there
is a momentum exchange between the system and the source of the field
via the Lorentz force, there is no energy exchange at $B=$~const.
Therefore, the term $PdV$ in Eq.~(\ref{eq:B27}) represents the work
done by the expanding system on the external confinement. To avoid
confusion, we note that the work done by the expanding system against
the Lorentz force actually goes to increase of the internal energy of
the system itself\cite{foot}.

\subsection{Fluid Dynamics Approach}

Consider an electron liquid in a uniform magnetic field and in the
presence of an electric potential, chemical potential, and temperature
all of which vary smoothly in space.  All the fields and currents in
the system are assumed to be either constant in time or varying at a
small frequency as discussed in Sec.~\ref{subsec:gencon}. We
concentrate on a small macroscopic element of the liquid with area
$\delta A$ in the interior of the sample, which we choose to be of a
size much less than the length scales of the fields and much larger
than the average inter-electron distance.

Then, setting $V=\phi$ in the RHS of (\ref{eq:B101}), and setting
$\partial \J / \partial t = 0$, as is appropriate for a
quasi-equilibrium situation, we find
\begin{equation}
n\nab\phi+\nab\cdot\hat\pi+
{e\over c}\J\times\B=0 ,
\label{eq:B105}
\end{equation}
where $n$, $\hat\pi$ and $\J$ are averaged over the element $\delta
A$.  If the induced current is small, then the correction to $\hat\pi$
arising from the current should be second order in $J$, and therefore
negligible.  Thus $\hat\pi$ may be replaced by its equilibrium value,
$P_{{\rm int}}\delta_{\alpha\beta}$, evaluated for the local values of
$\mu$ and $T$.  To first order, Eq.~(\ref{eq:B105}) becomes
\begin{equation}
\v\equiv{\J\over n}=
\frac{c}{eB}\hat\epsilon
\left(\nab\phi+{1\over n}\nab P_{{\rm int}}\right).
\label{eq:72}
\end{equation}
Apart from the additional term resulting from the pressure gradient,
the right-hand side of Eq.~(\ref{eq:72}) represents the classical
drift velocity in the crossed magnetic and electric fields.
Substituting Eq.~(\ref{eq:70}) into Eq.~(\ref{eq:72}) and using the
relation $n\nab\mu =\nab P - s\nab T$ which follows from
Eqs.~(\ref{eq:B43}) and (\ref{eq:B44}), we finally obtain
\begin{equation}
\J=n\v=\J_{{\rm tr}}-(c/e)\hat \epsilon\nab M,
\end{equation}
where $\J_{{\rm tr}}$ coincides with the right-hand side of
Eq.~(\ref{eq:69}), and the second term is the internal magnetization
current as defined by Eqs.~(\ref{eq:B3}) and~(\ref{eq:B7}).

We note that our arguments do not apply to an electron solid.  In a
solid, a non-uniform drift current will cause a shear deformation
which will increase until the stress forces suppress the local drift.
Equation~(\ref{eq:B105}) determines the force acting on an element of
a liquid and does not include the shear stress contribution when we
use (\ref{eq:B103}).  In addition, the pinning effects which arise in
the presence of even a weak disorder potential makes our consideration
completely inapplicable in the case of an electron solid.

\subsection{Derivation Using Onsager Symmetry}

We now study a homogeneous system, $\nabla T=\nabla\mu = 0$, which is
driven out of equilibrium by an external electric field ${\E} = (1/e)
\nab\phi$. Since we study coefficients of the bulk response, we can
assume that the electric field is uniform and the system itself is
infinite.  In addition to the original laboratory frame, we consider
the system in the ``primed'' reference frame moving at a velocity
\begin{equation}
\v=c[\E\times \B]/B^2=c\hat\epsilon\nab\phi/(eB),
\label{eq:73}
\end{equation}
in which the applied electric field vanishes.  Since the system is
homogeneous, and disorder potential is absent, Galilean invariance
requires that properties of the system in both reference frames are
the exactly same.  Due to the absence of electric field in the moving
frame, the system with respect to this frame is in equilibrium.  The
number and energy current densities in this frame are therefore zero,
$\J'=\J^{E'}=0$.  In the laboratory frame, the system moves as a whole
at a velocity $\v$.  Hence the number current density defined as the
average number of electrons passing through unit length per unit time
is given by
\begin{equation}
\J = n\v.
\label{eq:74}
\end{equation}
To determine the energy current density, we split the whole system in
two parts by an imaginary straight line perpendicular to the drift
velocity $\v$ and find out how much energy $\Delta E$ is transferred,
in the laboratory frame, from one part of the system to another in
time $\Delta t$.  The energy $\Delta E$ has two contributions, one
from the direct transfer of an element of the system across the line,
and another due to the work done by one part of the system on the
other part while moving
\begin{equation}
\Delta E = \epsilon\Delta x \Delta L + P_{{\rm int}} 
\Delta x \Delta L,
\label{eq:75}
\end{equation}
where $\Delta x = v \Delta t$ is displacement of the system in time
$\Delta t$, $\Delta L$ is the length of the line segment which we
consider, and $P_{{\rm int}}$ is the pressure in the interior of the
sample.  Substituting Eq.~(\ref{eq:70}) for the internal pressure into
(\ref{eq:75}), for the energy current density $J^E = \Delta E/ (\Delta
L \Delta t)$ we obtain
\begin{equation}
\J^E = (\epsilon + P_{{\rm int}} ) \v = 
(\epsilon + P-MB) \v .
\label{eq:76}
\end{equation}

One can also obtain Eq.~(\ref{eq:76}) directly from the microscopic
expression for the energy current, given by
Eqs.~(\ref{eq:B18})--(\ref{eq:21}).  Let us write $\v_i =\v'_i + \v$,
where $\v'_i$ is the velocity of particle $i$ in the frame moving with
velocity $\v$.  In the moving frame, the system is at local
equilibrium with an energy density $\epsilon_0$ and a stress tensor
$P_{{\rm int}}\delta_{ij}$.  Comparing (\ref{eq:B19})--(\ref{eq:21})
with (\ref{eq:B102}), we see that (\ref{eq:76}) is correct to first
order in $\v$.

Identifying Eqs.~(\ref{eq:74}) and~(\ref{eq:76}) with the expressions
for the bulk current response (\ref{eq:11}), (\ref{eq:12}), and
substituting $\v$ from Eq.~(\ref{eq:73}), we find the response
coefficients
\begin{eqnarray}
\hat L^{(1)} & = & -{nc\over eB}\hat\epsilon ,
\label{eq:77}
\\
\hat L^{(3)} & = & -{c(\epsilon+P-BM)\over
eB}\hat\epsilon.
\label{eq:78}
\end{eqnarray}
As one can see from the last formula, $\hat L^{(3)}$ is an odd
function of the magnetic field.  From this fact, and from the symmetry
relation $L_{\alpha\beta}^{(3)}(B) =L_{\beta\alpha}^{(2)}(-B)$
discussed in Sec.~\ref{subsec:onsager}, we have
\begin{equation}
\hat L^{(2)}=\hat L ^{(3)}=
-{c(\epsilon+P-BM)\over eB}
\hat\epsilon.
\label{eq:79}
\end{equation}
Substituting the obtained coefficients $\hat L^{(1)}$, $\hat L^{(2)}$,
and $\hat L^{(3)}$ into (\ref{eq:B34})--(\ref{eq:B36}), and using
(\ref{eq:B7}) and (\ref{eq:B44}), we find $\hat N^{(1)} = \hat
L^{(1)}$, and
\begin{equation}
\hat N^{(2)} = \hat N^{(3)} = -{sc\over eB} \hat\epsilon.
\label{eq:B301}
\end{equation}
Then using~(\ref{eq:B32}) we arrive at Eq.~(\ref{eq:69}) for the
transport number current.

\section{THERMOPOWER  MEASUREMENTS}

\subsection{Thermopower Measurements and Current Distributions}

We now turn to discuss the thermoelectric properties of real samples,
in which temperature gradients are maintained by the coupling of the
electron gas to the phonons of the substrate.  To use the results of
the previous sections, we shall assume that the sample is homogeneous,
and that the coupling to the substrate is sufficiently weak that the
corresponding thermal relaxation rate is much slower than the
microscopic relaxation rate of the electron gas $\tau_m^{-1}$, and the
response of the electron gas is well-described by the transport
properties of the isolated electron gas.  However, we assume that the
thermal coupling to the substrate is sufficiently strong that on a
macroscopic scale we may assume that the local temperature of the
two-dimensional electron gas is equal to the local temperature of the
substrate, and we need not impose the condition that $\nab\cdot \J^E =
0$ in the electron gas.  The substrate is assumed to be an electrical
insulator, however, so that $\nab\cdot\J = 0$ in the electron system.
We do not discuss the energy current in this section.

A convenient way in which to study the thermoelectric response of a
sample is through the thermopower.  A thermopower measurement involves
the application of a uniform temperature gradient $\nab T$ to a sample
which is disconnected from current leads.  Since there can be no
average electron current flow, an electrochemical potential gradient
$\nab\xi = \nab (\phi + \mu )$ develops.  The thermopower tensor $\hat
S$ is defined in terms of this potential gradient.

If the diagonal matrix elements of $\hat N^{(1)}$ and $\hat N^{(2)}$
are different from zero, it can be shown that the conditions $\nab
\cdot\J_{{\rm tr}} = 0$, with $\hn\cdot \J_{{\rm tr}} = 0$ at the
sample boundaries, together with Eq.~(\ref{eq:B32}), require that
$\J_{{\rm tr}} = 0$ everywhere.  This is the case only if
\begin{equation}
\nab \xi = {1\over e} \hat S \nab T
\label{eq:80}
\end{equation}
 everywhere in the sample, where
\begin{equation}
\hat S = -(eT)^{-1} [\hat N^{(1)} ]^{-1} [\hat
N^{(2)} ].
\label{eq:mynum3}
\end{equation}

In the special situation where $\hat N^{(1)}$ and $\hat N^{(2)}$ are
proportional to the antisymmetric tensor $\hat\epsilon$, as occurs for
instance in the case of zero impurities, the value of $\xi(\r )$ is
not properly determined by the conservation equations, together with
Eq.~(\ref{eq:B32}) and the boundary conditions.  For example, one may
add to $\xi$ any function $f(\r )$ which vanishes at the boundaries of
the sample, without affecting the values of $\nab \cdot\J_{{\rm tr}}$
in the interior or $\hn \cdot \J_{\rm tr}$ at the boundary.  The
average value of $\nab \xi$ is still given correctly by
Eq.~(\ref{eq:80}), however, for any solution of the equations, and the
value of $\xi$ at any point of the boundary will be the same as if
$\nab \xi$ were uniform in the sample.  Thus the experimentally
measured thermopower, in which the voltage drop is measured between
two points at the boundary, would still be given by Eq.~(\ref{eq:80})
in this case.

For the remainder of this section, we will focus on the thermopower of
systems in which the disorder potential is weak. In the limit of
vanishing disorder, one can use the response coefficients we have
derived in the previous section, and the thermopower tensor takes a
particularly simple form
\begin{equation}
S_{\alpha\beta}=- {s\over en}\delta_{\alpha\beta}.
\label{eq:82}
\end{equation}
Thus, the thermopower tensor is diagonal, with a magnitude given by
the entropy per particle, $s/n$, divided by the charge per particle,
$- e$.  This result is familiar for non-interacting
electrons\cite{obr65,jon,oji}.

Unfortunately it difficult to provide a general criterion for how weak
the impurity scattering must be in order that its effects on the
thermopower can be neglected and Eq.~(\ref{eq:82}) applies.  Rather,
the form of such a criterion depends on the nature of the low-lying
charged excitations of the system, which may be quite different at
different filling fractions (compare, for example, filling fractions
at which the system is compressible and incompressible in the
zero-temperature limit).  A necessary condition for Eq.~(\ref{eq:82})
to apply is that the impurity scattering is sufficiently weak that
both of the tensors $\hat{N}^{(1)}$ and $\hat{N}^{(2)}$ are almost
purely off-diagonal, such that the thermopower tensor itself is close
to diagonal.  In some circumstances this condition may not be
sufficient, as there may be corrections to the size of the diagonal
thermopower.  The form of such corrections depends on the specific
experimental conditions, and requires a specific calculation of the
effects of impurity scattering on the carriers.  In the following, we
will concentrate on thermopower of systems for which the impurity
scattering is sufficiently weak that Eq.~(\ref{eq:82}) applies.  In
the next subsection we will discuss the form of the corrections
that can arise due to impurity scattering for filling fractions close
to $\nu=1/2$ or $3/2$.

Although, under the conditions of the experiment, no net current
passes through the sample, in a quantizing magnetic field circulating
non-equilibrium currents are induced.  In the bulk, these are internal
magnetization currents, whose continuity at the edge is provided by
the boundary currents.  As we shall now show, these currents can be
very large, in the sense that the local current density in the
presence of both the temperature gradient and the compensating
electric field can be comparable to what it would have been in the
presence of only one of these fields.

Let us compare the average and the local current densities induced by
a uniform temperature gradient alone, $\nabla T =$~const, with
$\nabla\mu =\nabla\phi = 0$.  We consider a filling factor $\nu = 1/2$
which is an important and much studied example of a compressible
state.  The average current density is equal to the transport current
density (\ref{eq:69})
\begin{equation}
\jtr={cs\over eB}\hat\epsilon\nab T.
\label{eq:83}
\end{equation}
The portion of this current that flows in the bulk is given by
\begin{equation}
\J^{\bulk}=\jtr+\J^{\rm bulk}_{\rm mag}=
\left[\left. {cs\over eB}-{c\over e} {\partial s\over\partial B}
\right|_{\mu,T}\right]\hat\epsilon\nab T,
\label{eq:84}
\end{equation}
where we used Eq.~(\ref{eq:B22}) and the thermodynamic relations
(\ref{eq:B43})--(\ref{eq:B45}).

In a strong magnetic field, for which all electrons are restricted to
the lowest spin-polarized Landau level, one may express the entropy
per unit area in the form $s=n_0S_q[\nu,(e^2/\kappa\ell)/T]$, where
$n_0=eB/hc$ is the number density of flux quanta, $S_q$ is the entropy
per flux quantum, $\nu=n/n_0$ is the filling fraction, and we have
assumed a Coulomb force-law, for which the typical energy scale is set
by the magnetic length, $\ell\equiv\sqrt{\hbar c/eB}$.  Using
$\partial\ln(e^2/\kappa\ell)/\partial B=1/2$, one can write
\begin{equation}
\left.{\partial s\over
\partial B}\right|_{\mu,T}=
\left.{s\over B}-{n\over B}
{\partial S_q\over \partial\nu}+
{T\over 2B}{\partial s\over\partial T}
\right|_{n,B},
\label{eq:85}
\end{equation}
Now, at a filling fraction of one half, particle-hole symmetry
requires that $\partial S_q/\partial \nu=0$.  Hence Eq.~(\ref{eq:84})
may be rewritten
\begin{equation}
\J^{\bulk}(\nu=1/2)=-
\left.{cT\over 2eB}
{\partial s\over\partial T}\right|_{n,B}
\hat\epsilon\nab T.
\label{eq:86}
\end{equation}
We shall discuss two cases.

First, for an ideal non-interacting electron gas (for example
$\kappa\rightarrow\infty$), the entropy is independent of the
temperature at fixed filling fraction, so the above expression is
identically zero.  The temperature gradient induces no number current
in the bulk of the system, and all of the induced current flows around
the edge of the sample.

Secondly, for a system interacting by Coulomb forces, the entropy at
$\nu = 1/2$ is believed to be approximately linear in temperature
(with a logarithmic correction at very low temperature due to the
divergence of the effective mass of composite fermions)\cite{hlr}.
The bulk current induced by the temperature gradient is therefore
approximately $cs/ (2eB)\hat\epsilon\nab T$, which is {\it one
half} of the total current induced by the temperature gradient
(\ref{eq:83}); the remaining current flows around the edge of the
sample.

In both of the above cases the current driven by the temperature
gradient is found to be very inhomogeneous, with all or half of the
total current flowing on the edge of the sample.

In contrast, the distribution between edge and bulk of the current
driven by $\nab\xi$ in a thermopower experiment depends on the
apportionment between $\nab\mu$ and $\nab\phi$.  The relative
proportions of $\nab\xi$ coming from $\nab\mu$ and $\nab\phi$ depends
on the compressibility of the electron system and on the electrical
capacitance per unit area, i.e., on the distance to the nearest
conductivity surface.  In most practical cases, the contribution of
$\phi$ will be much larger than $\mu$, so that
$\nab\phi\approx\nab\xi$, and $\nab\mu\approx 0$.  In this case, the
compensating current driven by $\nab\xi$ will be uniformly spread over
the sample and a strong circulatory current is set up by the
combination of $\nab\xi$ and $\nab T$.  The non-equilibrium part of
the current density induced by the thermopower measurement has a form
shown schematically in Figure~2.
\begin{figure}
\inseps{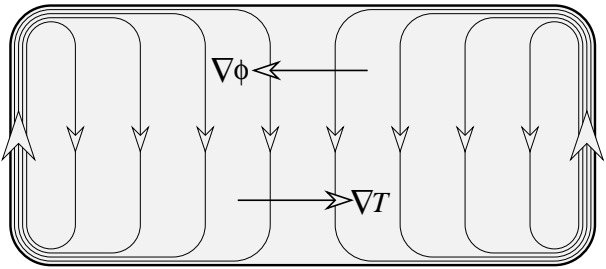}{0.34}
\vskip0.35cm
\caption{Schematic diagram of the distribution of the additional current
induced in a thermopower measurement in the quantum Hall regime. No
net current flows. However, a large fraction of the current induced by
the temperature gradient $\nabla T$ is at the edges of the sample,
whereas the compensating current induced by the electric field
$\nabla\phi$ is spread uniformly over the interior.}
\end{figure}

\subsection{Experimental Comparison}

Finally, we will compare our conclusions concerning the thermopower of
a sample in the limit of weak disorder with recent thermopower
measurements in the fractional quantum Hall regime.  At high
temperatures, the observed thermopower is dominated by the phonon-drag
contribution resulting from the momentum exchange between the system
and the phonons in the substrate.  Very low temperatures are required
before the intrinsic thermopower caused by the diffusion and drift in
the system itself can be observed.  It is only very recently that this
has been achieved in the fractional quantum Hall regime.  In
Ref.~\onlinecite{ying,bayot}, thermopower measurements on a hole gas
in this regime are reported, and it is found that at temperatures
below about 100 mK the intrinsic thermopower can be distinguished.
The crossover from a $T^3$ dependence of the thermopower at high
temperatures, to a linear temperature-dependence below 100 mK is
associated with the transition from phonon-drag-dominated to
diffusion-and-drift thermopower.  At such low temperatures the
signal-to-noise ratio is rather poor in thermoelectric measurements,
so it is difficult to resolve much structure related to the
incompressible states at fractional filling fractions.  One can,
however, distinguish dips in the diagonal thermopower, $S_{xx}$, at
$\nu = 1/3$, 2/5, 3/5, 2/3, consistent with the expectation that the
thermopower should vanish at these filling fractions for which the
entropy is exponentially small.

A more interesting issue is the behavior at even-denominator filling
fractions.  It is found that $S_{xx}$ exhibits a broad maximum at $\nu
= 1/2$ and, less clearly, at $\nu = 3/2$.  As explained in
Ref.~\onlinecite{ying} the absolute values of the thermopower are
inconsistent with a single particle picture, and electron-electron
interactions must be considered.

It has been argued that, at even denominator filling fractions, the
appropriate description of the electron system is in terms of a Fermi
liquid of weakly interacting composite fermions\cite{hlr}. We will use
this model in conjunction with equation (\ref{eq:69}) to calculate the
thermopower of a disorder-free system.  Let us first assume that
electrons are maximally spin-polarized.  Thus, we assume that at a
filling fraction of $\nu_i = i + 1/2$, the lowest $i$ (spin-split)
Landau levels are filled (and therefore contribute zero entropy), and
that the remaining half-filled Landau level may be represented by a
Fermi liquid of composite fermions with effective mass $m^*$.  The
entropy of the half-filled level is determined by the density of
states at the Fermi surface, so we obtain the thermopower of the
system to be
\begin{equation}
S_{\alpha\beta}= {\pi\over 6} {k_B^2m^*\over \hbar^2n q}
T\delta_{\alpha\beta},
\label{eq:87}
\end{equation}
where the particle charge $q$ is $-e$ for electrons and $+e$ for
holes.  Note that the same formula applies at all half-integer filling
fractions $\nu_i = i + 1/2$.  However, one must be careful to use the
appropriate value for the effective mass, which may vary for different
absolute magnetic fields and for different filling fractions $\nu_i$.

Corrections to Eq.~(\ref{eq:87}) due to impurity scattering may arise
if the scattering rate $\tau_i^{-1}$ of the composite fermions from
impurities is larger than the microscopic equilibration rate
$\tau_m^{-1}$ of the disorder-free composite-fermion system.  Since
this equilibration rate becomes very small for composite fermions
close to the Fermi surface, it may well be the case that in
experiments at low-temperature the impurity scattering rate is
sufficiently large, $\tau_i^{-1} \gg \tau_m^{-1}$, that corrections to
Eq.~(\ref{eq:87}) arise.  Arguments based on a Boltzmann transport
theory for the composite fermions suggest that, in this case, impurity
scattering will affect the thermopower (\ref{eq:87}) if the scattering
rate $\tau_i^{-1}$ is energy-dependent\cite{boltzmannfootnote}.
Specifically, if we consider a {\it} model of composite fermions with
conventional parabolic dispersion, $E\propto k^2$, and a transport
scattering rate that varies with energy as $E^{-p}$, then the effect
of impurity scattering is to multiply Eq.~(\ref{eq:87}) by a factor
$(1+p)$:
\begin{equation}
S_{\alpha\beta}\approx{\pi\over 6} (1+p){k_B^2m^*\over \hbar^2 n q}
T\delta_{\alpha\beta} ,
\label{eq:88}
\end{equation}
This has the same form as the conventional Mott formula for the
thermopower of a spinless two-dimensional electron gas in zero
magnetic field\cite{gal}.

A calculation\cite{khveshchenko} of the scattering of composite
fermions in modulation-doped quantum wells suggests that it is only
weakly energy-dependent, $p\simeq 0.13$, and the corrections to
(\ref{eq:87}) are small. In the following, we will compare the
experimental observations with Eq.~(\ref{eq:87}), bearing in mind that
a prefactor $(1+p)\simeq 1.13$ may arise due to impurity scattering.

Comparing Eq.~(\ref{eq:87}) with the measurements of $S_{xx}$ reported
by Ying {\it et al.}\cite{ying}, we find that, at a filling fraction
of $\nu = 1/2$ and at a magnetic field $B = 5.6\mbox{T}$, an effective
mass of $m^* = 1.3\pm 0.3 m_0$ is required for consistency, where
$m_0$ is the free-electron mass.  This value is a factor of 2 larger
than the value $m^*\approx 0.7 m_0$ obtained in Ref.~\onlinecite{ying}
from their own analysis of the data, which used the Mott formula
Eq.~(\ref{eq:88}) with an assumed value of $p\simeq 1$. However, our
value of the effective mass does not seem inconsistent with estimates
of $m^*$ based on other types of transport measurements.  For example,
a value of $m^* = 1.4 m_0$ at $\nu = 1/2$ is quoted in
Ref.~\onlinecite{ying} for a hole doped sample with a higher carrier
density, such that $\nu = 1/2$ occurred at $B = 13\mbox{T}$.  In an
ideal system, with no Landau-level mixing, zero layer thickness, and
no impurity corrections, the effective mass at $\nu = 1/2$ would be
expected to be proportional to $\sqrt B$, which would predict a value
$m^* \approx 0.9 m_0$ for the sample used for the thermopower
measurements, where $\nu = 1/2$ occurred at $5.6\mbox{T}$.  However,
it is not at all clear that this scaling should apply to the actual
samples.  (The observed value of the effective mass at $13\mbox{T}$ is
in any case considerably larger than one would expect based on
numerical studies of finite systems where Landau-level mixing, finite
thickness, and impurity effects are ignored.)

We note that the value $p\simeq 1$ assumed in Ref.~\onlinecite{ying}
was obtained from calculations of impurity scattering of electrons in
zero field, whereas calculations\cite{khveshchenko} for composite
fermions suggest a much smaller value $p\simeq 0.13$, as was mentioned
above.

The difference between our formula and the one used in
Ref.~\onlinecite{ying} is much more serious at $\nu = 3/2$.  In that
case Ying {\it et al.}\cite{ying} replace the particle density $n$ in
Eq.~(\ref{eq:88}) by the density of composite fermions, which is now 3
times smaller than the density of holes in the valence band.  We,
believe, however, that whether one uses Eq.~(\ref{eq:87}) or
Eq.~(\ref{eq:88}) the quantity $n$ should be the total number of
carriers, including those in any filled Landau levels.

Since the experimental thermopower reported in Ref.~\onlinecite{ying}
is larger at $\nu = 3/2$ than at $\nu = 1/2$ by a factor $\approx
1.4$, Ying {\it et al.} conclude that $m^*(3/2) \simeq 0.5 m^*(1/2)$,
which they view as evidence for the validity of a model of
spin-polarized composite fermions at $\nu=3/2$.  However, we would
conclude using Eq.~(\ref{eq:87}) or Eq.~(\ref{eq:88}) that $m^*
(3/2)\approx 1.4 m^* (1/2)$.  This is contrary to the expectation for
the ideal system that $m^* (3/2)$ should be smaller than $m^* (1/2)$,
due to the smaller value of $B$.

Unfortunately, we do not see any justification for the analysis used
by Ying {\it et al.} at $\nu=3/2$.  We believe that our starting
formula (\ref{eq:87}) is correct in the limit of small impurity
scattering, and that energy-dependent impurity scattering leads to an
additional prefactor $(1+p)$ that is close to unity.  Moreover, the
relation $m^*(3/2) / m^*(1/2) = S_{xx}(3/2) / S_{xx}(1/2)$ holds even
allowing such impurity scattering, provided the exponent $p$ is the
same at each filling fraction. (In the absence of any Landau level
mixing this would necessarily be the case if the magnetic length were
the same at each filling fraction. The change in magnetic length by a
factor of $\sqrt{3}$ that occurs between $\nu=1/2$ and $3/2$ at fixed
electron number density is unlikely to affect the exponent $p$.)

It therefore appears to us that the reported thermopower measurement
at $\nu = 3/2$ are not consistent with a simple model based on
spin-aligned composite fermions.  The failure of this model may be due
to the combined effects of increasing degree of Landau level coupling
and the smaller Zeeman energy expected at $\nu = 3/2$ as compared to
$\nu = 1/2$.  Alternatively, the effects of disorder may be quite
different at these two filling fractions.  Neglecting any significant
effects of disorder, however, and viewing the diagonal thermopower as
a measure of the entropy, one would conclude that the entropy at $\nu
= 3/2$ is larger than what one would expect from a model of maximally
spin-polarized composite fermions.  It may be that additional entropy
arises from the loss of spin-polarization.  A number of experiments
indicate that in typical electron doped GaAs samples, the electron
system is not maximally spin-polarized at $\nu=3/2$ even at
$T=0$\cite{unpol1,unpol2,unpol3,barrett}. It is not clear whether this
will also occur for hole-doped samples, where the Zeeman energy may be
more important due to the larger g-factor.  To gain better
understanding of the origin of the discrepancy at $\nu = 3/2$, it
would be interesting to investigate the dependence of the thermopower
on the extent of Landau level coupling (e.g., by studying $n$-type
samples, or $p$-type samples with different densities), and on the
Zeeman energy (by tilted field measurements).

\section{Conclusions}

We have discussed the linear response of a homogeneous, bounded
interacting electron gas in quantizing magnetic field. We studied the
number and energy currents which arise in response to the gradients in
electric and chemical potential and in temperature.  We derived
general expressions for the bulk and boundary currents in the presence
of mechanical and statistical fields.  In general, the boundary of the
sample can carry a finite fraction of the total current passing
through the sample.  The local response in the bulk may be described
as a sum of ``transport'' and ``internal magnetization''
contributions.  Internal magnetization currents do not contribute to
the net current, are always divergenceless, and cannot be revealed in
any standard transport experiments performed on either homogeneous or
inhomogeneous samples.  They can be detected only in special
contactless experiments resolving the local current distribution.  We
found that Onsager symmetry relations cannot, in general, be applied
directly to the local current densities in the bulk of the sample.
However, they do hold locally for the transport currents, and
therefore for the net currents passing through the sample.  We derived
expressions for three of the four response functions of an interacting
system in the limit of weak disorder in terms of equilibrium
properties of the system.  In particular, we showed that, in this
case, the thermopower tensor is diagonal and is proportional to the
entropy per particle.  Recent thermopower measurements on a
high-mobility sample show that this conclusion is consistent with a
model of a Fermi liquid of spin-polarized composite-fermions at $\nu =
1/2$.  However, for the observations to be consistent with this model
at $\nu = 3/2$, a very large effective mass is required.  An effective
mass of this size seems unlikely, and we suggest that the
spin-polarized composite-fermion state may not be a good description
of the system at that filling fraction.

\acknowledgements{The authors are grateful to M.~Shayegan 
for originally stimulating their interest in this problem, and for
helpful subsequent discussions.  We would also like to thank
P.N.~Butcher for helpful comments, and D.V.~Khveshchenko for providing
us with Ref.~\onlinecite{khveshchenko} prior to publication.  This
work has been supported in part by NSF grant DMR94--16910 and in part
by the NATO Science Fellowship Programme.}


\begin{thebibliography}{10}
\bibitem{foot} One can show this directly by considering the 
reference frame moving together with an element of the boundary
$\Delta\tau$, which is assumed to expand with a small velocity $\v$.
In this reference frame, there appears an electric field
${\E}=-(\v/c)\times\B$. The work done on the system by this electric
field in time $\Delta t$ is given by $\I\cdot\E\Delta t\Delta\tau$,
which is exactly the work done by the system against the Lorentz
force.
\bibitem{boltzmannfootnote}
One way of deriving Eq.~(\ref{eq:88}) is to calculate $\hat{N}^{(3)}$
and $\hat{N}^{(1)}$, by considering an infinite uniform sample with a
constant temperature and chemical potential and a constant electric
field $\E$, chosen so that the net current $J$ is in the
$y$-direction.  In a model where composite fermions are scattered by
impurities but not by each other, this means that the applied field
$E_x$ is cancelled by the Chern-Simons field, while $E_y$ is balanced
by momentum loss due to impurity scattering.  At $\nu=1/2$, where the
composite fermion current is equal to the total electric current, this
leads to the same Boltzmann equation, and the same ratio $J^Q/J$ as
one would have for non-interacting fermions in zero magnetic field,
for the given form of the impurity scattering.  Using
Eq.~(\ref{eq:mynum3}) and the Onsager relation
$\hat{N}^{(2)}=\hat{N}^{(3)}$, we see that the thermopower has the
same form in this case as for $B=0$.  At $\nu=3/2$, the value of
$\hat{N}^{(3)}$ is the same as at $\nu=1/2$ for the same density of
composite fermions and the same assumed impurity scattering, but the
value of $\hat{N}^{(1)}$ is three times larger. (We assume that the
Hall angle is still close to 90 degrees, so that $\hat{N}^{(1)}$ is
essentially given by the Hall conductance.)
\end{thebibliography}

\widetext

\end{document}